\documentclass[12pt]{article}
\usepackage{amsmath}
\usepackage{amssymb}
\usepackage{graphicx}
\begin{document}
\title{Dynamics of rogue waves on a multi-soliton background in a vector nonlinear Schr\"{o}dinger equation }
\author{GUI MU $^{\dag}$,\ ZHENYUN QIN  $^{\ddag}$\footnote{Corresponding author E-mail address: zyqin@fudan.edu.cn} AND ROGER GRIMSHAW $^{\S}$  \\
\it\small{{$^{\dag}$ College of Mathematics and Information Science, Qujing Normal University,
}}\\
\it\small{{ Qujing Yunnan  655011, P.R. China
}}\\
\it\small{{  $^{\ddag}$
School of Mathematical Sciences and Key Laboratory of Mathematics }}\\
\it\small{{for Nonlinear Science,
Fudan University, Shanghai 200433, P.R. China}}\\
\it\small{{ $^{\S}$
Department of Mathematical Sciences, Loughborough University, }}\\
\it\small{{ Loughborough, Leics. LE11 3TU, UK.}}
}
\date{}
\maketitle

\textbf{Abstract.} General higher order rogue waves of a vector nonlinear Schr\"{o}dinger equation (Manakov system) are derived using a Darboux-dressing transformation with an asymptotic expansion method. The $N$th order semi-rational solutions containing $3N$ free parameters are expressed in  separation of variables form.  These solutions exhibit rogue waves on a multisoliton background. They demonstrate that the structure  of rogue waves in this two-component system is richer than that in a one-component system. The study of our results would be of much importance in understanding and predicting rogue wave phenomena arising in nonlinear and complex systems, including optics, fluid dynamics, Bose-Einstein condensates and finance and so on.\\


\section{Introduction}

It is well known that many nonlinear wave equations of physical interest support solitons, which are localised waves  arising from a balance between dispersion and nonlinearity, and which can propagate steadily for a long time. Recently, over the last two decades, it has been recognised  that another class of solutions, namely breathers, are also of fundamental importance.  Breathers propagate steadily, and are localised in either time or space, while being periodic in either space or time.  Further, due to their localisation properties, breathers have been invoked as models of rogue waves, also called freak waves, which are large amplitude waves  which
apparently appear without warning, and then disappear without trace. While they have been most often found in the context of water waves \cite{kh}-\cite{sl}, they have also been found in other physical contexts such as optical fibres \cite{ki}-\cite{sol}.

The breather solutions of the  focusing nonlinear Schr\"{o}dinger equation (NLSE) have been widely invoked as  models of rogue waves,
see the references above and Akhmediev \cite{akh} for instance.
The NLSE equation is integrable \cite{za} and many kinds of exact solutions have been found.
 In particular, the Peregrine breather,\cite{pe}  the Akhmediev breather (AB) \cite{akh} and
 the Kuznetsov-Ma breathers (KM) \cite{ku}-\cite{ma} have been associated with rogue waves, as the potential outcome
 of the modulational instability of a plane wave. AB is periodic in space and localized in time, while KM is  periodic in time and localised in space.
 The Peregrine breather is especially considered as a rogue wave prototype because it is localised in both time and space,
 and so captures the fundamental features of rogue waves. Importantly it has a peak amplitude which is exactly three times
 the background. Also, it is the generic outcome of a wide class of modulated plane waves \cite{gr}.  When the spatial or temporal period is taken to be infinite, AB or KM  become Peregrine breather in the limit. Peregrine breather of NLSE has been observed experimentally in water wave tanks \cite{ch} and also in nonlinear fibre optics \cite{ki}-\cite{bi}.

The Peregrine breather is the lowest order item in a hierarchy of rational solutions for NLSE.
Since these higher-order forms may also be invoked as models for rogue waves, with higher amplitudes,
it is of interest to find explicit expressions for these higher-order rogue wave solutions.
For example, Akhmediev \cite{akha} presented some low-order rational solutions
using a Darboux transformation method. Ohta and Yang  \cite{oh} obtained  general high-order rogue waves
in terms of determinants using the Hirota bilinear method. Guo \cite{g1}-\cite{gu} constructed the $N$-order rogue wave solutions
using a generalized Darboux transformation. Dubard \emph{et al.} \cite{du} discussed the quasi-rational solutions via algebraic-geometry method.
As well as for the NLSE, rational solutions have also been explored for some other nonlinear wave equations in \cite{oh}, \cite{aaj},
\cite{he}-\cite{mu2}.

Here we shall extend a recent study of the integrable vector nonlinear Schr\"{o}dinger equations (VNLSE), or Manakov system, by
Baronio \emph{et al.} \cite{fa} This can be expressed in dimensionless form as,
\begin{equation}
\begin{split}
&i u_{1t}+u_{1xx}+2(|u_1|^2+|u_2|^2)u_1=0\,,\\
&i u_{2t}+u_{2xx}+2(|u_1|^2+|u_2|^2)u_2=0\,.
\end{split}
\label{1}
\end{equation}
Here  the subscripts $t,x$ stand for partial differentiation with resect to $t$, an evolution variable, and $x$, a spatial variable.
The dependent variables  $u_{1}, u_{2}$  represent  wave envelopes,  whose physical meaning
depends on the particular context. In particular  the system (\ref{1}) has applications
in nonlinear optics \cite{kau}-\cite{fo}
and in Bose-Einstein condensates \cite{bl}-\cite{ka}.
Also note  that  Eqs. (\ref{1}) correspond to the self-focusing (or anomalous dispersion) regime.
The fundamental vector rogue wave solutions have been recently reported by several authors \cite{fa},  \cite{bl}, \cite{gu0}.
Then the first-, second- and third-order rogue wave solutions of VNLSE were explicitly presented \cite{zhai}.
Here our aim is to find the general $N$-th order rogue wave solutions. We will construct  hierarchies of semi-rational solutions of
the VNLSE (\ref{1}),  using an asymptotic expansion method. Our obtained solutions are an extension of the results of
Baronio \emph{et al.} \cite{fa} to higher-order rogue wave solutions, and the pure rational solutions of our results
can be identified with the ones obtained by Zhai \emph{et al.} \cite{zhai} As the representations obtained are a combination of rational and exponential functions, some interesting structures can be observed, such as (1) the coexistence of higher-order rogue waves and
bright-dark multi-soliton solutions, and (2) the coexistence of multi-rogue waves and bright-dark multi-soliton solutions.
In general our obtained solutions indicate the complexity that can arise when rogue waves interact with solitons.

 The paper is organised as follows. In section 2, The Lax pair and Darboux-dressing transformation of VNLSE (\ref{1})
 are briefly reviewed. In section 3  the dynamic behaviour of a family of fundamental rogue wave solutions is examined.
 Then in section 4, we present some novel periodic  breathers. In section 5, we use an asymptotic expansion method
 to obtain $N$-th order  rogue wave solutions of VNLSE (\ref{1}). Then, in section 6, we give some examples to illustrate the
 a range of  dynamic behaviour of our obtained  rogue wave solutions. We conclude with a summary in section 7.

\section{Asymptotic expansion of Darboux-dressing transformation}

The VNLSE (\ref{1}) is integrable, and is a condition for the compatibility of the Lax pair,
\begin{equation}
\begin{split}
&\Psi_x={\bf U} \Psi = (i\lambda \sigma+Q)\Psi \,, \\
&\Psi_t={\bf V}\Psi = \left[2i\lambda^2\sigma+2\lambda Q+ i\sigma (Q^2-Q_x)\right]\Psi \,.
\end{split}
\label{2}
\end{equation}
Here $\Psi(x,t)$ is a $3\times 3$ matrix variable, $\lambda$ is the
complex spectral parameter, $\sigma=diag(1,-1,-1)$ is a constant diagonal matrix and $Q=Q(x,t)$ is the $3\times 3$ matrix variable
\begin{eqnarray}
Q=\left( \begin {array}{ccc} 0&-u_1^*&-u_2^*\\\noalign{\medskip}u_1&0&0\\\noalign{\medskip}u_2&0&0\end {array} \right),
\end{eqnarray}
It is straightforward  to check  the condition of compatibility between the two equation in (\ref{2})
\begin{eqnarray}
\Psi_{xt}=\Psi_{tx},
\end{eqnarray}
 leads directly to the VNLSE(\ref{1}).

A suitable Darboux-dressing transformation for the VNLSE (\ref{1}) is given by \cite{fa} \cite{deg}for instance,
\begin{eqnarray}
&&\Psi[1]=D\Psi,\quad D=I+\frac{(\lambda_1^{*}-\lambda_1)P}{\lambda-\lambda_1^{*}} \,,
\quad P=\frac{\psi_{0} \psi_{0}^{\dag}}{\psi_{0}^{\dag} \psi_{0}} \,, \label{2.4} \\
&&\left( \begin {array}{ccc} u_1\\\noalign{\medskip}u_2\end {array} \right)=
\left( \begin {array}{ccc} u_{10}\\\noalign{\medskip}u_{20}\end {array} \right)
+\frac{2 i (\lambda_1^{*}-\lambda_1)r^{*}}{|r|^2+|s_1|^2+|s_2|^2}\left( \begin {array}{ccc} s_1\\\noalign{\medskip}s_2\end {array} \right) \,. \label{2.5}
\end{eqnarray}
 Here, $I=\mathrm{diag}(1,1,1), \quad \psi_{0}=\Psi(x,t,\lambda_1)Z_0=(r(x,t),s_1(x,t),s_2(x,t))^T$ and $\Psi(x,t,\lambda_1)$ is the fundamental solution
 for  the Lax equations (\ref{2}) corresponding to $\lambda=\lambda_1$ and  for the seed solutions $\textbf{u}_0=(u_{10},u_{20})^T$ of the
 VNLSE (\ref{1}).   The constant parameter $\lambda_1$ is complex while $Z_0$ is an arbitrary nonzero complex 3-dimensional constant vector.
Next, it is useful to note that the Darboux-dressing transformation  (\ref{2.4}) can be replaced with the alternative form
\begin{eqnarray}\label{2.6}
&&\Psi[1]=T\Psi,\quad T=\lambda I+\Lambda \,, \quad
\Lambda=-\lambda_1^{*}I+(\lambda_1^{*}-\lambda_1)P \,, \quad P=\frac{\psi_{0} \psi_{0}^{\dag}}{\psi_{0}^{\dag} \psi_{0}} \,.
\end{eqnarray}
Since $T$ is also a Darboux transformation of NLSE, it follows that
\begin{eqnarray}\label{2.7}
&&T_x+T{\bf U}-{\bf U}_1T=0,\quad T_t+T{\bf V}-{\bf V}_1T=0,
\end{eqnarray}
The matrices ${\bf U}_1$ and ${\bf V}_1$ are obtained by replacing $Q$ with $Q_1$ in ${\bf U}$ and ${\bf V}$, respectively. Then inserting  (\ref{2.6}) into (\ref{2.7}) we find that
\begin{eqnarray}\label{2.8}
\begin{split}
&\Lambda_{x}+\Lambda Q_0-Q_1\Lambda=0,\\
&\Lambda_{t}+i\Lambda \sigma(Q_0^2-Q_{0x})-i\sigma(Q_1^2-Q_{1x})\Lambda=0,\\
\end{split}
\end{eqnarray}
\begin{eqnarray}\label{2.9}
\hspace{-2cm} \hbox{and} \quad Q_1=Q_0+i\Lambda\sigma-i\sigma\Lambda,
\end{eqnarray}

In general, for a Darboux transformation, the zero seed solution allows for the construction of a
hierarchy of multisoliton solutions , while a plane wave seed solution results in a hierarchy of breather-type solutions
related to modulation instability. However, here we note that
\begin{eqnarray}
&&T|_{\lambda=\lambda_1}\psi_{0}=(\lambda_1 I+\Lambda_1)\psi_{0}=0,
\end{eqnarray}
This means that the Darboux-dressing transformation (\ref{2.6}) cannot be iterated continuously for the same spectral parameter.
In order to eliminate this limitation, we introduce the following expansion theorem
which can then be used  to produce new solutions for the  same spectral parameter.\\

\noindent
\textbf{Theorem }\,\,\,  Let $\Psi[\lambda_1(1+\delta)]$ be a solution of the Lax pair system (\ref{2}) corresponding to the
spectral parameter $\lambda_1(1+\delta)$ and a seed solution $\textbf{u}_0=(u_1[0],u_2[0])^T$. If $\Psi[\lambda_1(1+\delta)]$
 has an expansion at $\lambda_1$
\begin{eqnarray}\label{a}
\Psi[\lambda_1(1+\delta)]=\Psi_{0}+\Psi_{1}\delta+\Psi_{2}\delta^2+\Psi_{3}\delta^3+\cdots,
\end{eqnarray}
\begin{eqnarray}\label{2.12}
\begin{split}
\hbox{then} \quad &\psi[n] = \left(r[n],s_1[n],s_2[n]\right)^T=\lambda_1\psi[n-1]+T[n]\Upsilon[n-1],\quad  n\geq 1,\\
&\psi[0] = \left(r[0],s_1[0],s_2[0]\right)^T=\Psi_{0},\\
&\Upsilon[n-1]=\psi[n-1](\Psi_j\rightarrow \Psi_{j+1}),\,\,\,j=0,1,2,\cdots
\end{split}
\end{eqnarray}
\begin{eqnarray*}
\hbox{where} \quad &&T[n]=\lambda_1 I+\Lambda[n],\quad \Lambda[n]=-\lambda_1^{*}+(\lambda_1^{*}-\lambda_1)P[n]\,,  \\
\hbox{and} \quad &&   P[n]=\frac{\psi[n-1] \psi[n-1]^{\dag}}{\psi[n-1]^{\dag} \psi[n-1]},
\end{eqnarray*}
are solutions of the Lax pair system (\ref{2}) corresponding to the same spectral parameter $\lambda_1$ and solution $\textbf{u}_{n}\equiv\left( \begin {array}{ccc} u_{1}[n]\\\noalign{\medskip}u_{2}[n]\end {array} \right)$
\begin{eqnarray*}
&&\textbf{u}_{n}=\left( \begin {array}{ccc} u_{1}[n-1]\\\noalign{\medskip}u_{2}[n-1]\end {array} \right)+\frac{2 \i (\lambda_1^{*}-\lambda_1)r[n-1]^{*}}{|r[n-1]|^2+|s_1[n-1]|^2+|s_2[n-1]|^2}\left( \begin {array}{ccc} s_1[n-1]\\\noalign{\medskip}s_2[n-1]\end {array} \right).
\end{eqnarray*}
\noindent
\textbf{Remark:} In the above Theorem, the denotation $\psi[n](\Psi_j\rightarrow \Psi_{j+1})$ means that $\Psi_j \,\,(j=1,2,\cdot\cdot\cdot n)$  are replaced correspondingly by $ \Psi_{j+1}$ but $T[j] \,\, (j=1,2,\cdot\cdot\cdot n)$  are left unchanged in $\psi[n]$. In addition, we introduce the denotation
\begin{eqnarray}
\Phi[N]=\Upsilon[N](\Psi_j\rightarrow \Psi_{j+1}),
\end{eqnarray}
The meaning of $\Upsilon[N](\Psi_j\rightarrow \Psi_{j+1})$ is same as  $\psi[n](\Psi_j\rightarrow \Psi_{j+1})$.
Concretely, let us illustrate these denotations by two simplest examples: for $n=0$, $\psi[0]=\Psi_0,\,\Upsilon[0]=\Psi_1,\,\Phi[0]=\Psi_2$. For $n=1$,
$\psi[1]=\lambda_1\psi[0]+T[1]\Upsilon[0]=\lambda_1\Psi_0+T[1]\Psi_1,\,\Upsilon[1]=\lambda_1\Upsilon[0]+T[1]\Phi[0]=\lambda_1\Psi_1+T[1]\Psi_2,\,
\Phi[1]=\lambda_1\Psi_2+T[1]\Psi_3$.


Generally, employing the iteration relation (\ref{2.12}) and with the help of the following expansion
\begin{eqnarray*}
&&P(\lambda)=(\lambda_1+T[n])(\lambda_1+T[n-1])\cdot\cdot\cdot(\lambda_1+T[1])\\
&&\quad\quad\,\,\,=f_0\lambda_1^n+f_1\lambda_1^{n-1}+f_2\lambda_1^{n-2}+\cdot\cdot\cdot+f_n= \sum_{j=0}^{n}g_j,
\end{eqnarray*}
with
\begin{eqnarray*}
&&g_j=\lambda_1^{n-j}f_j,\quad j=0,1,\cdot\cdot\cdot,n.\\
&&f_0=1,\quad f_1=T[1]+T[2]+\cdot\cdot\cdot+T[n],\quad \cdot\cdot\cdot,\\
&& f_n=T[n]T[n-1]\cdot\cdot\cdot T[1].
\end{eqnarray*}
Then $\psi[n]$ could be rewritten in a more explicit form
\begin{eqnarray}
\psi[n]=g_0\Psi_0+g_1\Psi_1+g_2\Psi_2+\cdot\cdot\cdot+g_n\Psi_n,
\end{eqnarray}
It is natural to obtain
\begin{eqnarray}\label{m}
\Upsilon[n]=g_0\Psi_1+g_1\Psi_2+g_2\Psi_3+\cdot\cdot\cdot+g_n\Psi_{n+1},
\end{eqnarray}
and
\begin{eqnarray}\label{n}
\Phi[n]=g_0\Psi_2+g_1\Psi_3+g_2\Psi_4+\cdot\cdot\cdot+g_n\Psi_{n+2},
\end{eqnarray}
Based on these preparations, we turn to the proof of Theorem.
\noindent\\
\textbf{Proof:}\quad Substituting (\ref{a}) into (\ref{2}) and equation to zero, the coefficients of $\delta^j\quad (j=0,1,\cdot\cdot\cdot)$, we arrive at
\begin{eqnarray}\label{2.14}
\begin{split}
&\Psi_{0x}=U_0\Psi_{0},\\
&\Psi_{jx}=i\lambda_1\sigma\Psi_{j-1}+U_0\Psi_{j}, (j\geq 1)
\end{split}
\end{eqnarray}
and
\begin{eqnarray}\label{2.15}
\begin{split}
&\Psi_{0t}=V_0\Psi_{0},\\
&\Psi_{1t}=V_0\Psi_{1}+
\left(4i\lambda_1^2\sigma+2\lambda_1 Q_0\right)\Psi_{0},\\
&\Psi_{jt}=V_0\Psi_{j}+
\left(4i\lambda_1^2\sigma+2\lambda_1 Q_0\right)\Psi_{j-1}\\
&\quad\quad+2\i\lambda_1^2\sigma\Psi_{j-2}, (j\geq 2)
\end{split}
\end{eqnarray}
with
\begin{eqnarray*}
U_0=i\lambda_1\sigma+Q_0,\quad V_0=2i\lambda_1^2\sigma+2\lambda_1 Q_0+ i\sigma (Q_0^2-Q_{0x}),
\end{eqnarray*}
From the first equation of (\ref{2.14}) and (\ref{2.15}), we conclude that $\psi[0]=(r[0],s_1[0],s_2[0])^T=\Psi_0$ is  a solution of the the linear systems (\ref{2}) with $\textbf{u}_0$ and $\lambda=\lambda_1$. Thus, we can construct a  Darboux transformation as in (\ref{2.6}) above
\begin{eqnarray}
&&\Psi[1]=T[1]\Psi,\quad T[1]=\lambda I +\Lambda[1],\\ \nonumber
&&\Lambda[1]=-\lambda_1^{*} I +(\lambda_1^*-\lambda_1)P[1],\quad P[1]=\frac{\psi[0]\psi[0]^{\dag}}{\psi[0]^{\dag}\psi[0]},
\end{eqnarray}
\begin{eqnarray}\label{2.17}
\hbox{and then} \quad T|_{\lambda=\lambda_1}\psi[0]=T[1]\Psi_{0}=0.
\end{eqnarray}
Meanwhile, the corresponding solution of the VNLSE is given by
\begin{eqnarray}
&&\textbf{u}_{1}\equiv\left( \begin {array}{ccc} u_{1}[1]\\\noalign{\medskip}u_{2}[1]\end {array} \right)=\left( \begin {array}{ccc} u_{1}[0]\\\noalign{\medskip}u_{2}[0]\end {array} \right)+\frac{2i (\lambda_1^{*}-\lambda_1)r[0]^{*}}{|r[0]|^2+|s_1[0]|^2+|s_2[0]|^2}\left( \begin {array}{ccc} s_1[0]\\\noalign{\medskip}s_2[0]\end {array} \right),
\end{eqnarray}
Thus we have shown that the Theorem holds for $n=0$.

Next we use mathematical induction for integers $n \ge 1$.
Assume the Theorem holds for $n\leq N$, that is,
\begin{eqnarray}
\label{2.18} &&\psi[n]_{x}=(i\lambda_1\sigma+ Q_n)\psi[n],\\
\label{2.19} &&\psi[n]_{t}=\left[2i\lambda^2\sigma+2\lambda Q_{n}+ i\sigma (Q_{n}^2-Q_{nx})\right]\psi[n],
\end{eqnarray}
\begin{eqnarray*}
\hbox{with} \quad \psi[n]=\lambda_1\psi[n-1]+T[n]\Upsilon[n-1],\,\,\Upsilon[n-1]=\psi[n-1](\Psi_j\rightarrow \Psi_{j+1}) \,.
\end{eqnarray*}
Employing the following properties
\begin{eqnarray}\label{2.20}
&&\psi[N-1]_{x}=(i\lambda_1\sigma+ Q_{N-1})\psi[N-1],\\
&&\label{2.21} T[N]_{x}=\Lambda[N]_{x}=Q_N\Lambda[N]-\Lambda[N]Q_{N-1},\quad T[N]\psi[N-1]=0,\\
&&Q_N=Q_{N-1}+i\Lambda[N]\sigma-i\sigma\Lambda[N],\label{2.22}
\end{eqnarray}
the left side of (\ref{2.18}) is given by
\begin{eqnarray*}
&&(\lambda_1\psi[N-1]+T[N]\Upsilon[N-1])_x\\
&&=\lambda_1\psi[N-1]_{x}+T[N]_{x}\Upsilon[N-1]+T[N]\Upsilon[N-1]_{x},\\
&&=\lambda_1(i\lambda_1\sigma+ Q_{N-1})\psi[N-1]+(Q_{N}\Lambda_{N}-\Lambda[N]Q_{N-1})\Upsilon[N-1]+T[N]\Upsilon[N-1]_{x},
\end{eqnarray*}
Comparing this  with the right side of (\ref{2.18}), we get that
\begin{eqnarray}\label{2.23}
&&T[N]\Upsilon[N-1]_{x}=(\lambda_1(i\lambda_1\sigma+ Q_{N})+i\lambda_1\sigma\Lambda[N]+\Lambda[N]Q_{N-1})\Upsilon[N-1] \nonumber\\
&&\quad\quad\quad\quad\quad\quad\quad\quad+\lambda_1(Q_N-Q_{N-1})\nonumber\psi[N-1]\\
&&=(i\lambda_1^2\sigma+ \lambda_1(Q_{N-1}+i\Lambda[N]\sigma-i\sigma\Lambda[N])+i\lambda_1\sigma\Lambda[N]+\Lambda[N]Q_{N-1})\Upsilon[N-1] \nonumber\\
&&\quad+\lambda_1(i\Lambda[N]\sigma-i\sigma\Lambda[N])\psi[N-1] \nonumber\\
&&=T[N](i\lambda_1\sigma+Q_{N-1})\Upsilon[N-1]+i\lambda_1T[N]\sigma\psi[N-1],
\end{eqnarray}
Factoring out the factor $T[N]$ in (\ref{2.23}), we arrive at
\begin{eqnarray}\label{2.28}
\Upsilon[N-1]_{x}=(i\lambda_1\sigma+Q_{N-1})\Upsilon[N-1]+i\lambda_1\sigma\psi[N-1],
\end{eqnarray}
Furthermore, from (\ref{m}), (\ref{n}) and (\ref{2.14}),  we have
\begin{eqnarray}\label{y}
&&\nonumber \Upsilon[n]_{x}=\sum_{j=0}^{n}(g_{jx}\Psi_{j+1}+g_{j}\Psi_{j+1,x})\\
&&\quad\quad\,\,\,=\sum_{j=0}^{n}(i\lambda_1\sigma\Psi_j+(g_{jx}+U_0)\Psi_{j+1}),
\end{eqnarray}
and
\begin{eqnarray}\label{z}
&&\nonumber\Phi[n]_{x}=\sum_{j=0}^{n}(g_{jx}\Psi_{j+2}+g_{j}\Psi_{j+2,x})\\
&&\quad\quad\,\,\,=\sum_{j=0}^{n}(i\lambda_1\sigma\Psi_{j+1}+(g_{jx}+U_0)\Psi_{j+2}),
\end{eqnarray}
For an arbitrary nonnegative integer $n$, by a comparison (\ref{y}) with (\ref{z}), it indicate that $\Phi[n]_{x}$ can be obtained directly  after replacing $\Psi_j$ by $\Psi_{j+1}$ in $\Upsilon[n]_{x}$. As a result, from (\ref{2.28}), we immediately get
\begin{eqnarray}\label{g}
\Phi[N-1]_{x}=(i\lambda_1\sigma+Q_{N-1})\Phi[N-1]+i\lambda_1\sigma\Upsilon[N-1],
\end{eqnarray}
Now, with the help of (\ref{2.28}) and (\ref{g}), it is derived that
\begin{eqnarray}
&&\nonumber\Upsilon[N]_{x}=(\lambda_1\Upsilon[N-1]+T[N]\Phi[N-1])_x\\
&&\nonumber=\lambda_1\Upsilon[N-1]_x+T[N]_x\Phi[N-1]+T[N]\Phi[N-1]_x\\
&&\nonumber=\lambda_1((i\lambda_1\sigma+Q_{N-1})\Upsilon[N-1]+i\lambda_1\sigma\psi[N-1])+(Q_N\Lambda[N]-\Lambda[N]Q_{N-1})\Phi[N-1]\\
&&\nonumber\quad+T[N]((i\lambda_1\sigma+Q_{N-1})\Phi[N-1]+i\lambda_1\sigma\Upsilon[N-1])\\
&&\nonumber=(i\lambda_1^2\sigma+\lambda_1Q_{N-1}+i\lambda_1T[N]\sigma)\Upsilon[N-1]+i\lambda_1^2\sigma\psi[N-1]\\
&&\nonumber\quad+(Q_N\Lambda[N]-\Lambda[N]Q_{N-1}+i\lambda_1T[N]\sigma+T[N]Q_{N-1})\Phi[N-1]\\
&&\nonumber=(i\lambda_1^2\sigma+\lambda_1Q_{N}+i\lambda_1\sigma\Lambda[N]+i\lambda_1^2\sigma)\Upsilon[N-1]+i\lambda_1^2\sigma\psi[N-1]\\
&&\nonumber\quad+(Q_N\Lambda[N]+i\lambda_1T[N]\sigma+\lambda_1Q_{N-1})\Phi[N-1]\\
&&\nonumber=(i\lambda_1^2\sigma+\lambda_1Q_{N})\Upsilon[N-1]+i\lambda_1\sigma\psi[N]\\
&&\nonumber\quad+(Q_N\Lambda[N]+i\lambda_1^2\sigma+\lambda_1Q_{N}+i\lambda_1\sigma\Lambda[N])\Phi[N-1]\\
&&\nonumber=(i\lambda_1\sigma+Q_{N})(\lambda_1\Upsilon[N-1]+T[N]\Phi[N-1])+i\lambda_1\sigma\psi[N]\\
&&=(i\lambda_1\sigma+Q_{N})\Upsilon[N]+i\lambda_1\sigma\psi[N],\label{h}
\end{eqnarray}
The assumption (\ref{2.18}) allows us to construct the next step of the Darboux dressing transformation $T[N+1]$ which satisfies
\begin{eqnarray}\label{k}
\begin{split}
&T[N+1]_{x}=\Lambda[N+1]_{x}=Q_{N+1}\Lambda[N+1]-\Lambda[N+1]Q_{N},\\
&T[N+1]\psi[N]=(\lambda_1 I+\Lambda[N+1])\psi[N]=0,\\
&Q_{N+1}=Q_{N}+i\Lambda[N+1]\sigma-i\sigma\Lambda[N+1],
\end{split}
\end{eqnarray}
Therefore, using (\ref{h}) and (\ref{k}), it follows that
\begin{eqnarray}\label{2.24}
&&\nonumber T[N+1]\Upsilon[N]_{x}=(\lambda_1 I+\Lambda[N+1])((i\lambda_1\sigma+Q_{N})\Upsilon[N]+i\lambda_1\sigma\psi[N])\\
\nonumber&&=(\lambda_1 I+\Lambda[N+1])(i\lambda_1\sigma+Q_{N})\Upsilon[N]+i\lambda_1(\lambda_1 I+\Lambda[N+1])\sigma\psi[N]\\
\nonumber&&=(i\lambda_1^2\sigma+\lambda_1 Q_{N}+i\lambda_1\Lambda[N+1]\sigma+\Lambda[N+1]Q_{N})\Upsilon[N]\\
\nonumber&&\quad+i\lambda_1(-\sigma\Lambda[N+1]+\Lambda[N+1]\sigma)\psi[N]\\
\nonumber&&=(\lambda_1(i\lambda_1\sigma+ Q_{N+1})+i\lambda_1\sigma\Lambda[N+1]+\Lambda[N+1]Q_{N})\Upsilon[N]\\
&&\quad+\lambda_1(Q_{N+1}-Q_{N})\psi[N],\label{2.34}
\end{eqnarray}
Next, we need to show that
\begin{eqnarray}
&&\psi[N+1]_{x}=(i\lambda_1\sigma+Q_{N+1})\psi[N+1] \,.
\end{eqnarray}
Indeed, using (\ref{2.18})-(\ref{2.24}), we get that
\begin{eqnarray*}
&&\psi[N+1]_{x}=(\lambda_1\psi[N]+T[N+1]\Upsilon[N])_x\\
&&=\lambda_1\psi[N]_{x}+\Lambda[N+1]_{x}\Upsilon[N]+T[N+1]\Upsilon[N]_{x}\\
&&=\lambda_1(i\lambda_1\sigma+Q_N)\psi[N]+(Q_{N+1}\Lambda[N+1]-\Lambda[N+1]Q_{N})\Upsilon[N]\\
&&\quad+(\lambda_1(i\lambda_1\sigma+ Q_{N+1})+i\lambda_1\sigma\Lambda[N+1]+\Lambda[N+1]Q_{N})\Upsilon[N]
 \\ && \quad +\lambda_1(Q_{N+1}-Q_{N})\psi[N]\\
&&=\lambda_1(i\lambda_1\sigma+Q_{N+1})\psi[N]+(i\lambda_1\sigma+Q_{N+1})(\lambda_1+\Lambda[N+1])\Upsilon[N]\\
&&=(i\lambda_1\sigma+Q_{N+1})(\lambda_1\psi[N]+T[N+1]\Upsilon[N])\\
&&=(i\lambda_1\sigma+Q_{N+1})\psi[N+1],
\end{eqnarray*}
Similarly, for the temporal flow, we can show that
\begin{eqnarray}
&&\psi[N+1]_{t}=\left[2i\lambda^2\sigma+2\lambda Q_{N+1}+ i\sigma (Q_{N+1}^2-Q_{N+1,x})\right]\psi[N+1] \,.
\end{eqnarray}
This completes the proof by induction. \hfill $\square $

In addition, by a slight adjustment of the above proof of our Theorem,
it will provide a justifiable way to prove the Theorem 2 in Guo \emph{et al}. \cite{g1}.

\section{Exact breather solutions of the  Lax pair system}

Before using the expansion Theorem to construct higher-order rogue wave solutions, we use the
Darboux-dressing transformation (\ref{2}) to construct a new class of breather solutions.
It is readily shown that  the VNLSE (\ref{1}) admit the  plane wave background solution,
\begin{eqnarray}\label{3.1}
u_{10}=a_1 e^{i\varphi},\,\,\,
u_{20}=a_2e^{i\varphi},
\end{eqnarray}
\begin{eqnarray*}
\hbox{where} \quad \varphi=kx+(2\omega^2-k^2)t,\quad  \omega=\sqrt{a_1^2+a_2^2},
\end{eqnarray*}
where $a_1$ and $a_2$ are arbitrary parameters which, without loss
of generality, are taken as real, and $k$ is a wavenumber.

The corresponding solution of the Lax system (\ref{2})  is sought in the form
\begin{eqnarray}\label{3.2}
\Psi =\left(r(x,t), s_1(x,t), s_2(x,t) \right)^T=AFGz \,,
\end{eqnarray}
\begin{eqnarray}
&&F=\exp(i\Theta x),\quad G=\exp(i\Omega t),\quad A=\mathrm{diag}(1,e^{i\varphi},e^{\i\varphi}),
\end{eqnarray}
and $z=(\mu_1,\mu_2,\mu_3)^T$ is an arbitrary complex vector, and $v = \Psi z$.
Here, it is required that the constant matrices $\Theta$ and $\Omega$ satisfy
\begin{eqnarray}\label{6}
[\Theta,\Omega]=\Theta\Omega-\Omega\Theta=0 \,.
\end{eqnarray}
Inserting (\ref{3.2}) into (\ref{2}) yields
\begin{eqnarray}\label{7}
\begin{split}
&A_x+i A\Theta-{\bf U}A=0,
&A_t+i A\Omega-{\bf V}A=0,
\end{split}
\end{eqnarray}
Solving the conditions (\ref{6}) and (\ref{7}),  we obtain
\begin{eqnarray*}
&&\Theta=\left( \begin {array}{ccc} \lambda &i a_1&i a_2\\\noalign{\medskip}-i a_1\quad &-\lambda-k&0\\\noalign{\medskip}-i a_2&0&-\lambda-k\end {array} \right),\\
&&\Omega=\Theta^2+2\lambda\Theta-\lambda^2-2\omega^2,
\end{eqnarray*}
Then the exponential  matrices $F$ in (\ref{3.2}) can be written as
\begin{eqnarray}\label{3.6}
F=\frac{1}{2\tau}\exp(-i kx/2)\left( \begin {array}{ccc} \eta_1\quad&\eta_2\quad&\eta_3\\\noalign{\medskip}-\eta_2\quad&\eta_4\quad&\eta_5\\\noalign{\medskip}-\eta_3\quad&\eta_5\quad&\eta_6\end {array} \right),
\end{eqnarray}
where
\begin{eqnarray*}
&&\eta_1=2\tau\cosh(\tau x)+i \mu\sinh(\tau x),\quad \eta_2=-2a_1\sinh(\tau x),\quad \eta_3=-2a_2\sinh(\tau x),\\
&&\eta_4=a_1^2\omega^{-2}\left[2\tau\cosh(\tau x)-i \mu\sinh(\tau x)\right]+2a_2^2\tau\omega^{-2}e^{-i\mu x/2},\\
&&\eta_5=a_1a_2\omega^{-2}\left[2\tau\cosh(\tau x)-i \mu\sinh(\tau x)-2\tau e^{-i\mu x/2}\right],\\
&&\eta_6=a_2^2\omega^{-2}\left[2\tau\cosh(\tau x)-i \mu\sinh(\tau x)\right]+2a_1^2\tau\omega^{-2}e^{-i\mu x/2},\\
&&\tau=\frac{i}{2}\sqrt{\mu^2+4\omega^2},\quad \mu=2\lambda+k \,.
\end{eqnarray*}
Similarly the exponential  matrices $G$ in (\ref{3.2}) can be written as
 \begin{eqnarray}\label{3.7}
G=\frac{1}{2\xi}\exp({i k^2 t/2-i \omega^2 t})\left( \begin {array}{ccc} \kappa_1\quad&\kappa_2\quad&\kappa_3\\\noalign{\medskip}-\kappa_2\quad&\kappa_4\quad&\kappa_5\\\noalign{\medskip}-\kappa_3\quad&\kappa_5\quad&\kappa_6\end {array} \right),
\end{eqnarray}
where
\begin{eqnarray*}
&&\kappa_1=2\xi\cosh(\xi t)+i \mu\nu\sinh(\xi t),\quad \kappa_2=-2a_1\nu\sinh(\xi t),\quad \kappa_3=-2a_2\nu\sinh(\xi t),\\
&&\kappa_4=a_1^2\omega^{-2}\left[2\xi\cosh(\xi t)-i \mu\nu\sinh(\xi t)\right]+2a_2^2\xi\omega^{-2}e^{-i(\omega^2t+\mu\nu t/2)},\\
&&\kappa_5=a_1a_2\omega^{-2}\left[2\xi\cosh(\xi t)-i \mu\nu\sinh(\xi t)-2\xi e^{-i(\omega^2t+\mu\nu t/2)}\right],\\
&&\kappa_6=a_2^2\omega^{-2}\left[2\xi\cosh(\xi t)-i \mu\nu\sinh(\xi t)\right]+2a_1^2\xi\omega^{-2}e^{-i(\omega^2t+\mu\nu t/2)},\\
&&\xi=\nu \tau,\quad \nu=2\lambda-k.
\end{eqnarray*}

\begin{figure*}
  \centering
  \includegraphics[width=5in]{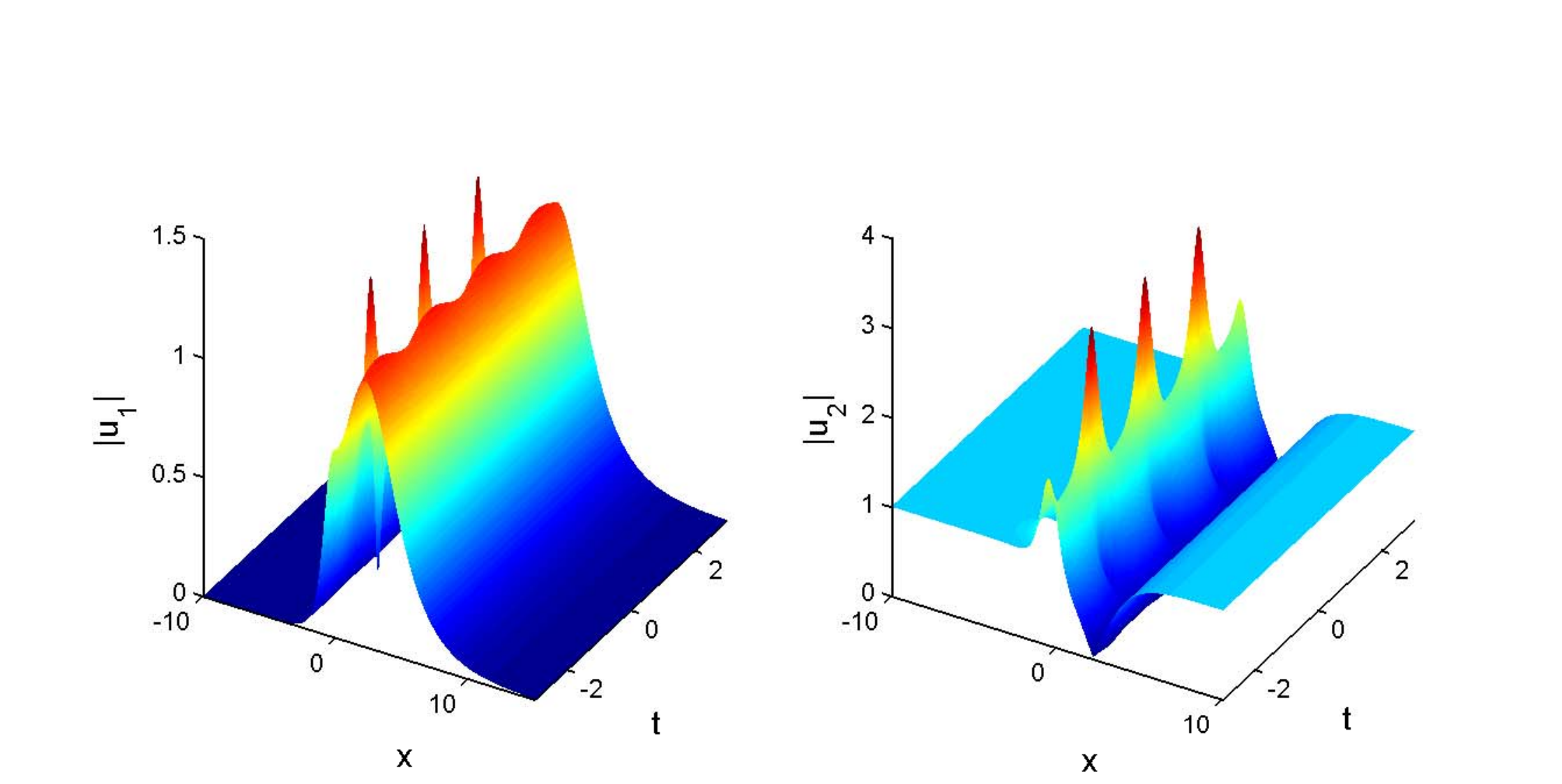}\\
  \caption{ (Color online) Bright-dark temporally periodic breather of the VNLSE (\ref{1})
  for parameters $a_1=0, a_2=1, k=0, \lambda=\frac{5}{4}i, \mu_1=0, \mu_2=1, \mu_3=2$.}
  \label{fig:1}
\end{figure*}

\begin{figure*}
  \centering
  \includegraphics[width=5in]{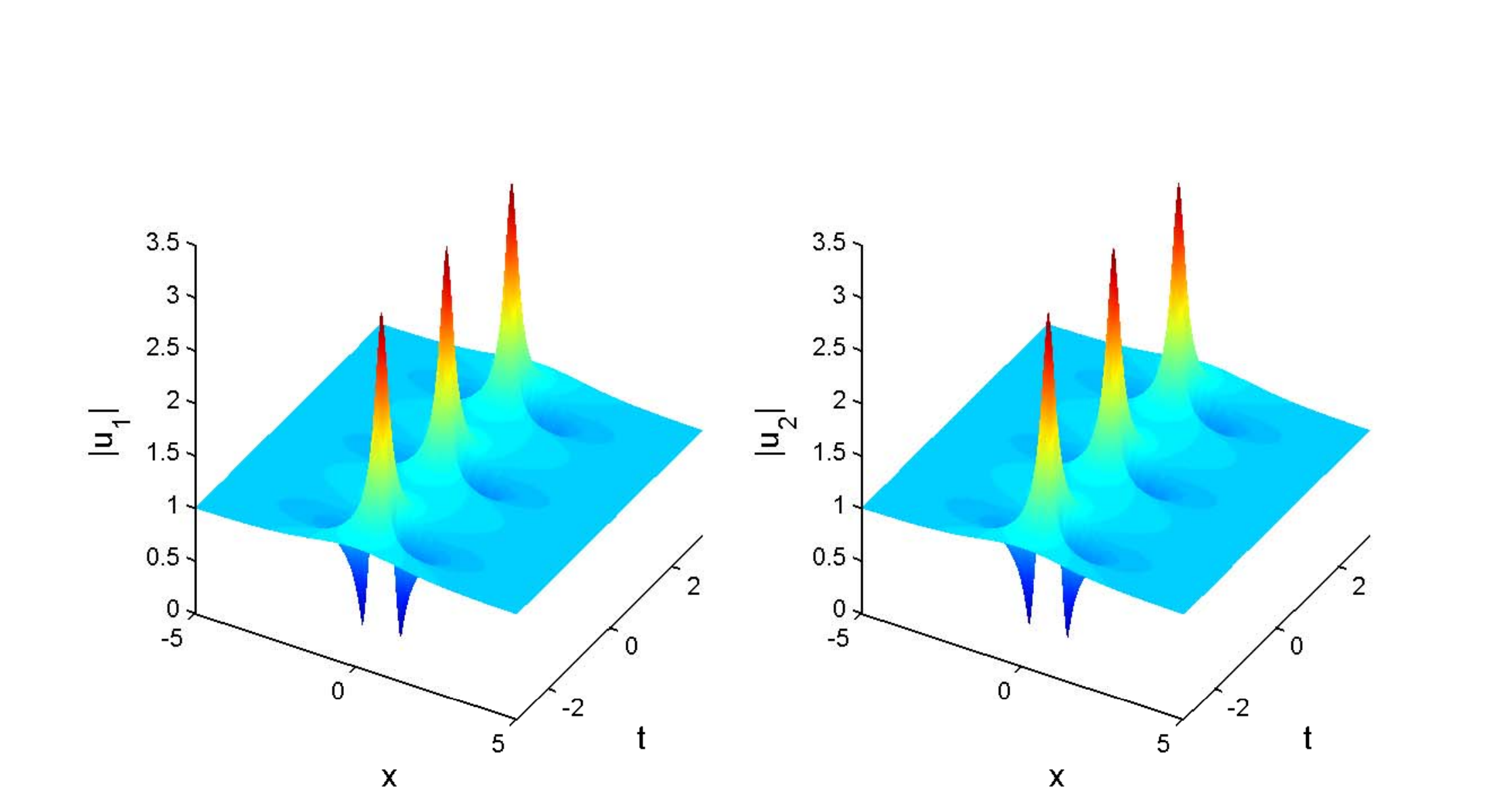}\\
  \caption{ (Color online) A typical temporal periodic breather of the VNLSE (\ref{1})
  for parameters $a_1=1, a_2=1, k=0, \lambda=\frac{3}{2}i, \mu_1=1, \mu_2=1, \mu_3=1$.}
  \label{fig:2}
\end{figure*}

\begin{figure*}
  \centering
  \includegraphics[width=5.5in]{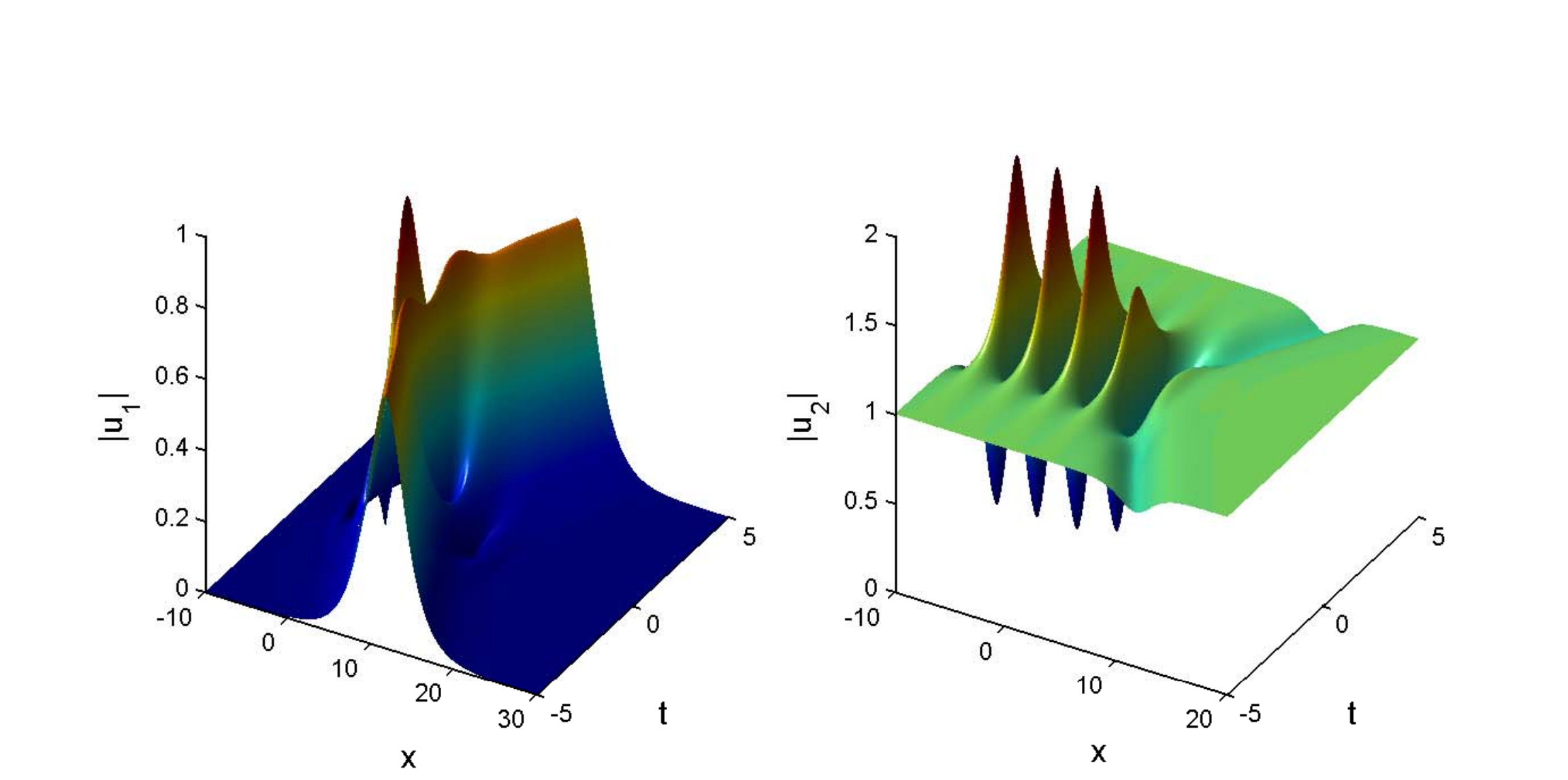}\\
  \caption{ (Color online) Bright-dark spatially periodic breather of the VNLSE (\ref{1})
  for parameters $a_1=0, a_2=1, \lambda=\frac{i}{2}, \mu_1=1, \mu_2=1, \mu_3=5$.}
  \label{fig:3}
\end{figure*}

\begin{figure*}
  \centering
  \includegraphics[width=5.5in]{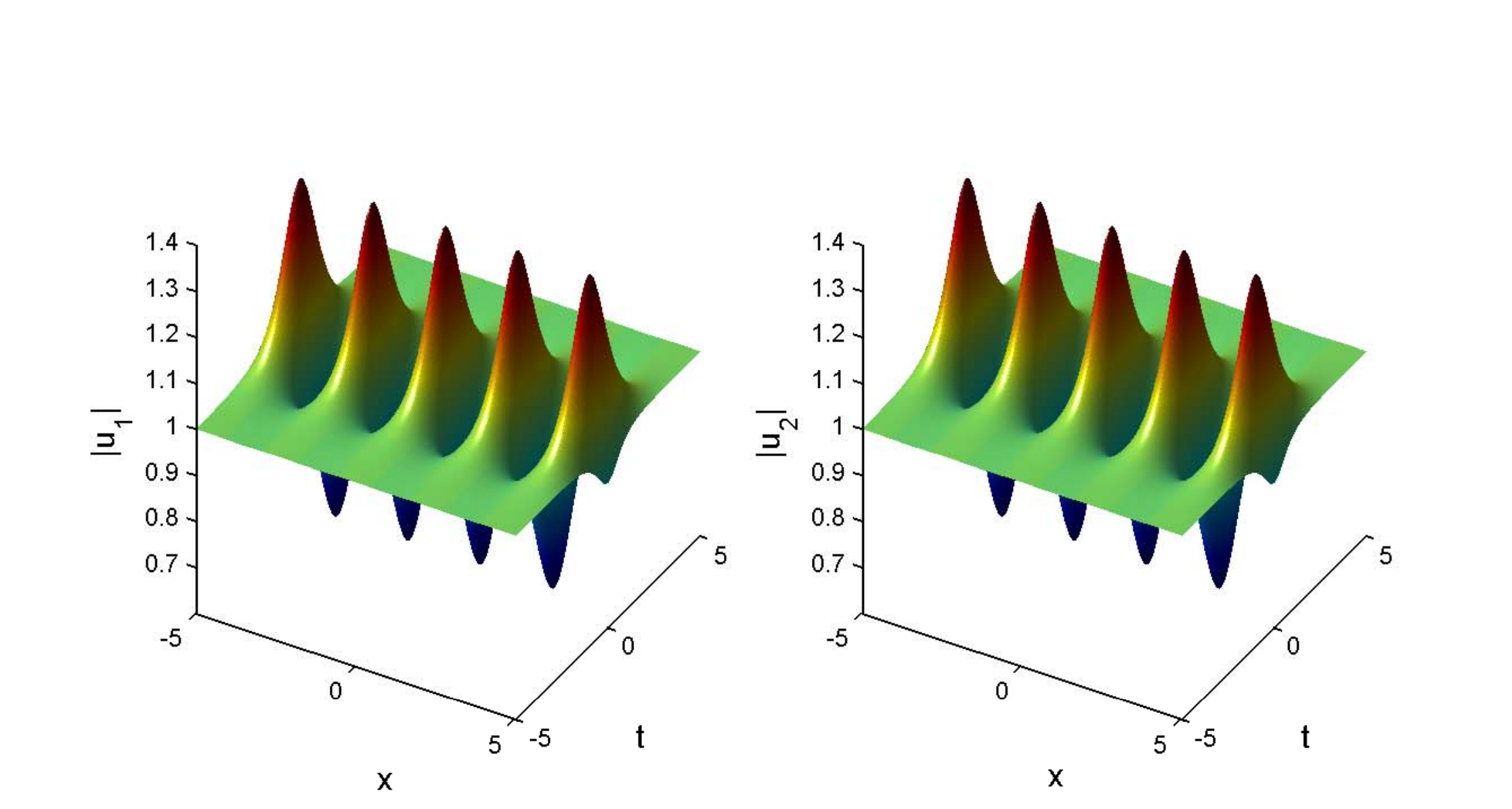}\\
  \caption{ (Color online) Typical spatially periodic breather of the VNLSE (\ref{1}0
  for parameters $a_1=1, a_2=1, \lambda=\frac{i}{4}, \mu_1=1, \mu_2=1, \mu_3=1$.}
  \label{fig:4}
\end{figure*}

Next, using the plane wave (\ref{3.1}) as the seed solution in the Darboux transformation (2.5) we can construct a
new exact solution of the VNLSE (\ref{1}) composed of hyperbolic functions and exponential functions.
 Since their full expressions are cumbersome, we omit them here.
 In particular, we note that  when $\lambda=i h,\, -2\omega<h<2\omega$, the solutions become spatially periodic.
 On the other hand,  when $\lambda=i h,\, h>2\omega$ or $h<-2\omega$, the solutions become temporally periodic.
 Figures \ref{fig:1} and \ref{fig:2}  show  two different temporally periodic breathers
 while Figures \ref{fig:3} and \ref{fig:4}  show two spatially periodic breathers.
 Due  to the exponential functions in these solutions,
 the dynamic features of these breathers are different from the cases studied by Forest  \cite{fo}.

Further,the exponential functions in these  solutions can be eliminated if the components of the vector $Z_0$ satisfy the condition
\begin{eqnarray}
a_1\mu_3-a_2\mu_2=0,
\end{eqnarray}
Under this constraint, the dynamic behavior will then be similar to those of  AB and KM in the NLSE
(see Figures \ref{fig:2} and \ref{fig:4}). It is well known that the Peregrine breather  is  a limiting case of
AB and KM when the period of either solution becomes infinite, and the same process could be employed here.\\


\begin{figure*}
  \centering
    \includegraphics[width=5in]{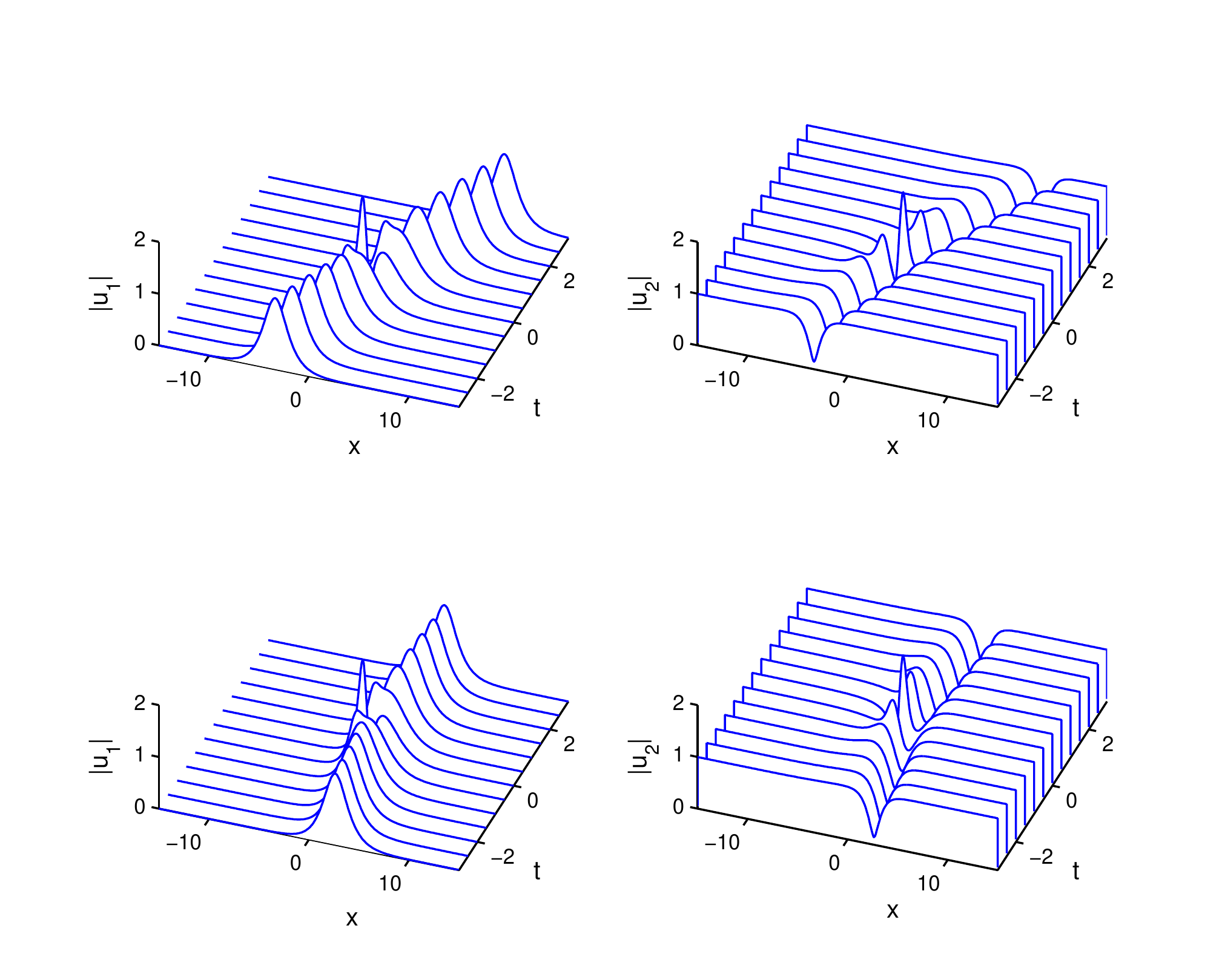}\\
  \caption{ (Color online)  Bright-dark rogue waves are emerging in the process of soliton propagation. The parameters are $a_1=0, a_2=1, \mu_1=1, \mu_2=1, \mu_3=\frac{1}{2}$. For wavenumber $k$, the top plane corresponds to $k=2$ while the bottom corresponds to $k=0$. }
  \label{fig:5}
\end{figure*}

When $\lambda=-\frac{k}{2}\pm i\omega$, the exponential $\exp(i\Lambda x+i\Omega t)$ becomes a
combination of exponential and polynomial functions of $x$
and $t$. In particular, taking $\lambda=-\frac{k}{2}+i\omega$, we arrive at a family of the solution with a  wavenumber $k$
\begin{eqnarray}\label{3.9}
\left( \begin {array}{ccc} u_1\\\noalign{\medskip}u_2\end {array} \right)=e^{i\theta}\left( \begin {array}{ccc} a_1\\\noalign{\medskip}a_2\end {array} \right)+\frac{4\omega \zeta^{*}}{|\zeta|^2+|z_1|^2+|z_2|^2}\left( \begin {array}{ccc} z_1\\\noalign{\medskip}z_2\end {array} \right),
\end{eqnarray}
where
\begin{eqnarray*}
&&  (\zeta\,\,\,z_1\,\,\,z_2)^T=AFGZ_0,\quad
Z_0=(\mu_1\,\,\,\mu_2\,\,\,\mu_3)^T,\\
&&  F=\left(\begin {array}{ccc} 1-\omega x&-a_1x&-a_2x\\\noalign{\medskip}a_1x&\omega^{-2}(a_1^2\varpi+a_2^2e^{\omega x})\quad &a_1a_2\omega^{-2}(\varpi-e^{\omega x})\\\noalign{\medskip}a_2x&a_1a_2\omega^{-2}(\varpi-e^{\omega x})\quad &\omega^{-2}(a_2^2\varpi+a_1^2e^{\omega x})\end {array} \right),
\end{eqnarray*}
\begin{eqnarray*}
&&G=\left(\begin {array}{ccc} 1+\i \mu_0\nu_0 t/2 &-a_1\nu_0  t&-a_2\nu_0  t\\\noalign{\medskip} a_1\nu_0  t&  a_1^2\zeta_1+a_2^2\zeta_2\quad &a_1a_2(\zeta_1-\zeta_2)\\\noalign{\medskip} a_2\nu_0  t&a_1a_2(\zeta_1-\zeta_2)\quad &a_2^2\zeta_1+a_1^2\zeta_2 \end {array} \right),
\end{eqnarray*}
with $\varpi=1+\omega x, \zeta_1=\omega^{-2}(1-i \mu_0\nu_0 t/2), \zeta_2=\omega^{-2} e^{-i(\omega^2t+\mu_0\nu_0 t/2)}$
and $\mu_0=2i \omega, \nu_0=2i \omega-2k$.
This is plotted in Figure \ref{fig:5} and illustrates how rogue waves can
occur in the  propagation of a bright-dark soliton. Note  that the bright-dark soliton with a non-zero wavenumber has
a slight change in velocities and phases when compared to the case with a zero wavenumber.
Similar dynamic behaviour occurs for the other values of $a_1, a_2$ and $Z$, but the corresponding figures are omitted here.

\section{Derivation of $N$th-order rogue waves solutions}

In this section, we derive $N$th-order rogue waves of the VNLSE (\ref{1} ) using our expansion theorem of section 2.
For simplicity, we consider only the case of zero wavenumber $k=0$. Thus we let $\lambda=i \omega (1+\delta) (0<\delta<1)$ in (\ref{3.2}).
Using Taylor series expansions expansions for the trigonometric and exponential functions,
the matrix $F$ in (\ref{3.6}) can be expanded around  $\delta=0$ as
\begin{eqnarray}\label{5.2}
F|_{\lambda=i\omega(1+\delta)}=\sum_{n=0}^{\infty}F_n\delta^n,
\end{eqnarray}
where
\begin{eqnarray*}
F_n=\left( \begin {array}{ccc} \vartheta_{n}&-a_1\omega^{-1}\beta_{n}&-a_2\omega^{-1}\beta_{n}\\\noalign{\medskip}a_1\omega^{-1}\beta_{n}\quad &\omega^{-2}
\left(a_1^2\Pi_{n}+a_2^2e^{\omega x}A_n\right)\quad &a_1a_2\omega^{-2}\left(\Pi_{n}-e^{\omega x}A_n\right)\\\noalign{\medskip}a_2\omega^{-1}\beta_{n}&a_1a_2\omega^{-2}\left(\Pi_{n}-e^{\omega x}A_n\right)&\omega^{-2}\left(a_2^2\Pi_{n}+a_1^2 e^{\omega x}A_n\right)\end {array} \right),
\end{eqnarray*}
with
\begin{eqnarray*}
&&\vartheta_{n}=\alpha_{n}-\beta_{n}-\beta_{n-1},\quad \Pi_{n}=\alpha_{n}+\beta_{n}+\beta_{n-1},\quad A_m=\frac{\omega^{m}x^{m}}{m!},\\
&&\alpha_{n}=\sum_{l=0}^{\lfloor n/2\rfloor}C_{n-l}^{l}2^{n-2l}A_{2(n-l)},\,\, \beta_{n}=\sum_{l=0}^{\lfloor n/2\rfloor}C_{n-l}^{l}2^{n-2l}A_{2(n-l)+1},
\,\, C_{n}^m=\frac{n!}{m!(n-m)!} \,.
\end{eqnarray*}
Here $n$ and $m$ are nonnegative integers with $n\geq m$.  Similarly, the matrix  $G$ in (\ref{3.7}) has the expansion
\begin{eqnarray}\label{5.3}
G|_{\lambda=i\omega(1+\delta)}=e^{-i\omega^2 t}\sum_{n=0}^{\infty}G_n\delta^n,
\end{eqnarray}
where
\begin{eqnarray*}
G_n=\left( \begin {array}{ccc} \sigma_{n}&-a_1\omega^{-1}\gamma_{n}&-a_2\omega^{-1}\gamma_{n}\\\noalign{\medskip}a_1\omega^{-1}\gamma_{n}\quad &\omega^{-2}
\left(a_1^2\chi_{n}+a_2^2e^{i\omega^2 t}\rho_{n}\right)\quad &a_1a_2\omega^{-2}\left(\chi_{n}-e^{i\omega^2 t}\rho_{n}\right)\\\noalign{\medskip}a_2\omega^{-1}\gamma_{n}&a_1a_2\omega^{-2}\left(\chi_{n}-e^{i\omega^2 t}\rho_{n}\right)&\omega^{-2}\left(a_2^2\chi_{n}+a_1^2 e^{i\omega^2 t}\rho_{n}\right)\end {array} \right),
\end{eqnarray*}
with
\begin{eqnarray*}
&&\sigma_{n}=\gamma_{n}-\theta_{n}-\theta_{n-1},\quad \chi_{n}=\gamma_{n}+\theta_{n}+\theta_{n-1},\\
&&\gamma_{n}=\sum_{l=0}^{\lfloor 3n/4\rfloor}\sum_{m=0}^{l}(-1)^{n-l}C_{n-l}^{m}C_{2(n-l)}^{l-m}2^{n-l-m}B_{2(n-l)}, \\
&&\theta_{n}=\i\sum_{l=0}^{\lfloor(3n+1)/4\rfloor}\sum_{m=0}^{l}(-1)^{n-l}C_{n-l}^{m}C_{2(n-l)+1}^{l-m}2^{n-l-m}B_{2(n-l)+1}, \\
&&\rho_{n}=\sum_{l=0}^{\lfloor n/2\rfloor}C_{n-l}^{l}i^{n-l}2^{n-2l}B_n, \quad B_m=\frac{2^m\omega^{2m}t^m}{m!}.
\end{eqnarray*}
On the above formulas, $l$ should be a nonnegative integer. Next, let us assume $Z_0$  be an arbitrary polynomial function of $\delta$ given by
\begin{eqnarray}\label{5.4}
Z_0(\delta)=\sum_{k=0}^{n}W_k\delta^k,
\end{eqnarray}
where $W_k=(\mu_{1k}\,\,\, \mu_{2k}\,\,\, \mu_{3k})^T$ are arbitrary  constant vectors.
Thus,
\begin{eqnarray*}
&& \Psi|_{\lambda=i\omega(1+\delta)}=\sum_{n=0}^{\infty}\Psi_n\delta^n, \quad \Psi_n=e^{-i\omega^2 t}AH_n,
\quad  H_n=\sum_{k=0}^{n}\sum_{j=0}^{n}F_kG_{j} W_{n-k-j},
\end{eqnarray*}
Then taking $\lambda_1=i\omega$ in the Theorem yields the $N$-rogue waves solutions of the VNLSE (\ref{1}).
Since the matrixes $F_i$ are independent of the variable $t$ while the matrixes $G_i$ are independent of the variable $x$,
the $N$th rogue wave solutions for VNLSE are expressed explicitly in a separation of variable form.

Moreover, if the components of the vector $W_k$ satisfy the conditions
\begin{eqnarray}
a_1\mu_{3k}-a_2\mu_{2k}=0,\,\,\,k=1\cdot\cdot\cdot N,
\end{eqnarray}
then solution (17) can be reduced to an $N$th-order pure rational solution. In particular, when $N=1,2,3$,
the solutions reduce to the cases considered by Zhai \emph{et al}. \cite{zhai}.

\section{Dynamics of rogue wave solutions}

In this section, we discuss the dynamics of these high-order rogue-wave solutions.
First setting $N =1$, the first-order rogue wave is produced,
\begin{eqnarray}
\left( \begin {array}{ccc} u_{11}\\\noalign{\medskip}u_{21}\end {array} \right)=e^{2i\omega^2t}\left( \begin {array}{ccc} a_1\\\noalign{\medskip}a_2\end {array} \right)+\frac{4\omega r[0]^{*}}{|r[0]|^2+|s_{1}[0]|^2+|s_{2}[0]|^2}\left( \begin {array}{ccc} s_{1}[0]\\\noalign{\medskip}s_{2}[0]\end {array} \right),
\end{eqnarray}
\begin{eqnarray*}
&&\hbox{where} \quad \left(r[0],s_1[0],s_2[0]\right)^T \equiv \psi[0]=\Psi_0=e^{-i\omega^2 t}AF_0G_0W_0,\\
&&F_0=\left(\begin {array}{ccc} 1-\omega x&-a_1x&-a_2x\\\noalign{\medskip}a_1x&\omega^{-2}(a_1^2\varpi+a_2^2e^{\omega x})\quad &a_1a_2\omega^{-2}(\varpi-e^{\omega x})\\\noalign{\medskip}a_2x&a_1a_2\omega^{-2}(\varpi-e^{\omega x})\quad &\omega^{-2}(a_2^2\varpi+a_1^2e^{\omega x})\end {array} \right),\\
&&G_0=\left(\begin {array}{ccc} -2i\omega^2 t+1&-2i a_1\omega t&-2i a_2\omega t\\\noalign{\medskip}2i a_1\omega t&a_2^2\omega^{-2} e^{i\omega^2 t}+ a_1^2\Gamma \quad &a_1a_2(\Gamma-\omega^{-2} e^{i\omega^2 t})\\\noalign{\medskip}2a_2\omega i t&a_1a_2(\Gamma-\omega^{-2} e^{i\omega^2 t})\quad  & a_1^2\omega^{-2} e^{i\omega^2 t}+ a_2^2\Gamma \end {array} \right),
\end{eqnarray*}
with the following notation: $\varpi=1+\omega x, \Gamma=\omega^{-2}+2\i t$.
This agrees with (\ref{3.9}) in the case of $k=0$ and has been discussed in detail in the work  \cite{fa}.

Taking $N =2$, the second-order rogue wave  is produced,
\begin{eqnarray}\label{6.2}
\left( \begin {array}{ccc} u_{12}\\\noalign{\medskip}u_{22}\end {array} \right)=\left( \begin {array}{ccc} u_{11}\\\noalign{\medskip}u_{21}\end {array} \right)+\frac{4\omega \,\,r[1]^{*}}{|r[1]|^2+|s_{1}[1]|^2+|s_{2}[1]|^2}\left( \begin {array}{ccc} s_{1}[1]\\\noalign{\medskip}s_{2}[1]\end {array} \right),
\end{eqnarray}
\begin{eqnarray*}
&&\hbox{where} \quad (r[1]\,\,\, s_1[1]\,\,\,s_2[1])^T\equiv \psi[1]=2\i \omega\left[(I-P[1])\Psi_1+\frac{1}{2}\psi_{0}\right],\\
&&P[1]=\frac{\psi[0]\psi[0]^{\dag}}{\psi[0]^{\dag}\psi[0]},\quad \Psi_1=e^{-i\omega^2 t}A\left[(F_0G_1+F_1G_0)W_0+F_0G_0W_1\right],
\end{eqnarray*}
The matrixes $F_1$ and $G_1$ are given by
\begin{eqnarray*}
F_1=\left( \begin {array}{ccc} -\omega x(1-\omega x+\frac{\omega^2 x^2}{3})&-\frac{1}{3}a_1\omega^2x^3&-\frac{1}{3}a_2\omega^2x^3\\\noalign{\medskip}\frac{1}{3}a_1\omega^2x^3 &a_1^2H+a_2^2\omega^{-1}xe^{\omega x}& a_1a_2\left(H-\omega^{-1}xe^{\omega x}\right)\\\noalign{\medskip}\frac{1}{3}a_2\omega^2x^3& a_1a_2\left(H-\omega^{-1}xe^{\omega x}\right)&a_2^2H+a_1^2\omega^{-1}xe^{\omega x}\end {array} \right),
\end{eqnarray*}
\begin{eqnarray*}
G_1=\left( \begin {array}{ccc} -4\omega^2t(i+\omega^2 t-\frac{2}{3}i\omega^4 t^2)&\frac{2}{3}i a_1\omega t(-3+4\omega^4t^2)&\frac{2}{3}i a_2\omega t(-3+4\omega^4t^2)\\\noalign{\medskip}\frac{2}{3}i a_1\omega t(3-4\omega^4t^2) &a_1^2J+4i a_2^2te^{\i\omega^2 t}& a_1a_2\left(J-4i te^{i\omega^2 t}\right)\\\noalign{\medskip}\frac{2}{3}i a_2\omega t(3-4\omega^4t^2)& a_1a_2\left(J-4i te^{i\omega^2 t}\right)&a_2^2J+4i a_1^2te^{i\omega^2 t}\end {array} \right),
\end{eqnarray*}
with the following notation: $H=\omega^{-1}x+x^2+\frac{\omega x^3}{3}$ and $J=4t(i-\omega^2 t-\frac{2}{3}i\omega^4 t^2)$.


If the conditions $a_1\mu_{30}-a_2\mu_{20}=0$ and $a_1\mu_{31}-a_2\mu_{21}=0$ are imposed,
the expression (\ref{6.2}) yields a trival generalisation of the rational Peregrine solution.
This is the situation considered in the work  \cite{zhai}.
In this case, the component $u_1(x,t)$ is merely proportional to $u_2(x,t)$.
Here, we are concerned with other parameter values.

\begin{figure*}
  \centering
  \includegraphics[width=5in]{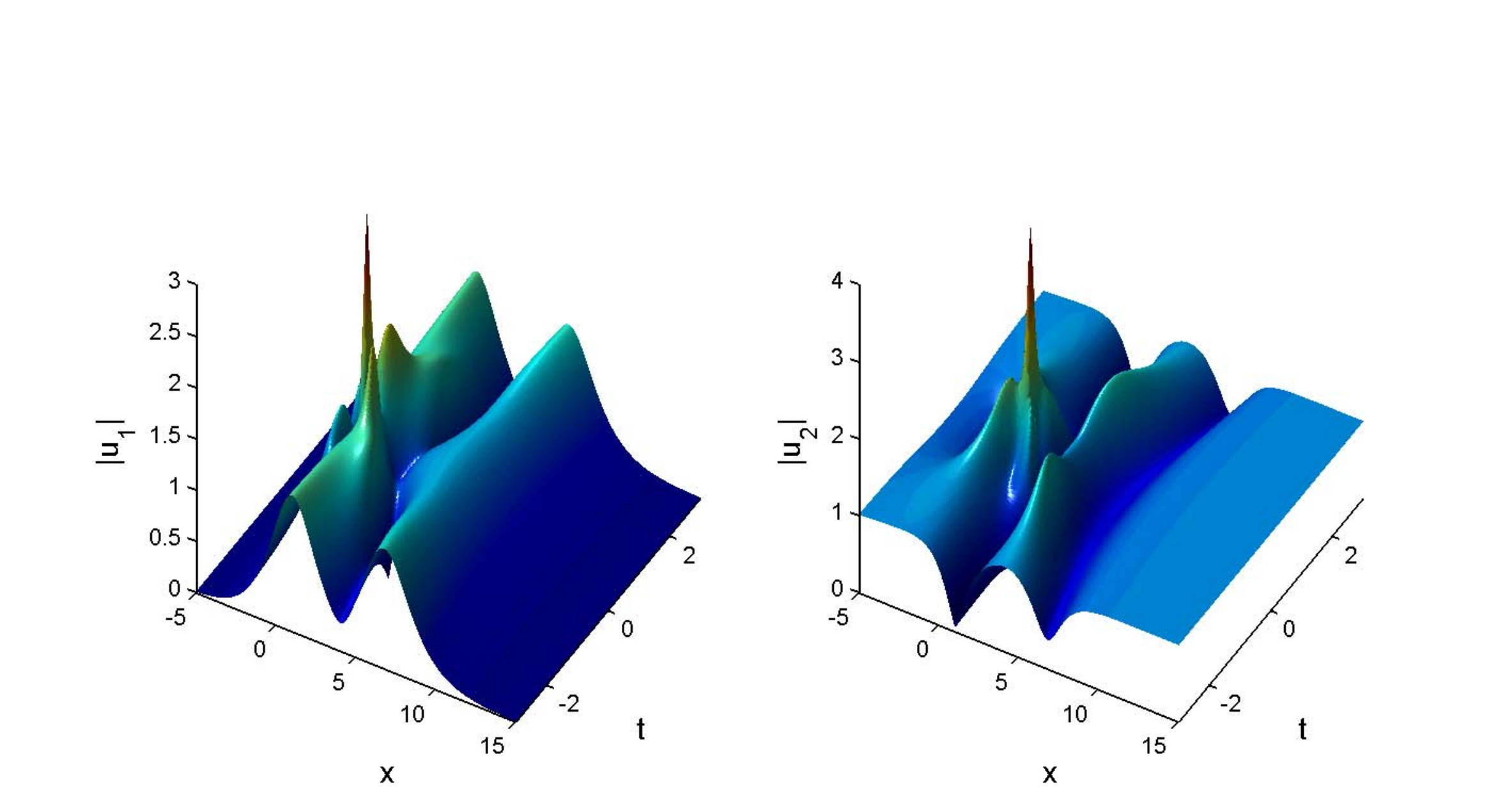}\\
  \caption{ (Color online) Second order rogue waves emerging in the propagation of two bright-dark solitons. The parameters are $a_1=0, a_2=1, \mu_{10}=1, \mu_{20}=1, \mu_{30}=1, \mu_{11}=0, \mu_{21}=0$ and $\mu_{31}=0$.}
  \label{fig:6}
\end{figure*}

\begin{figure*}
  \centering
  \includegraphics[width=5in]{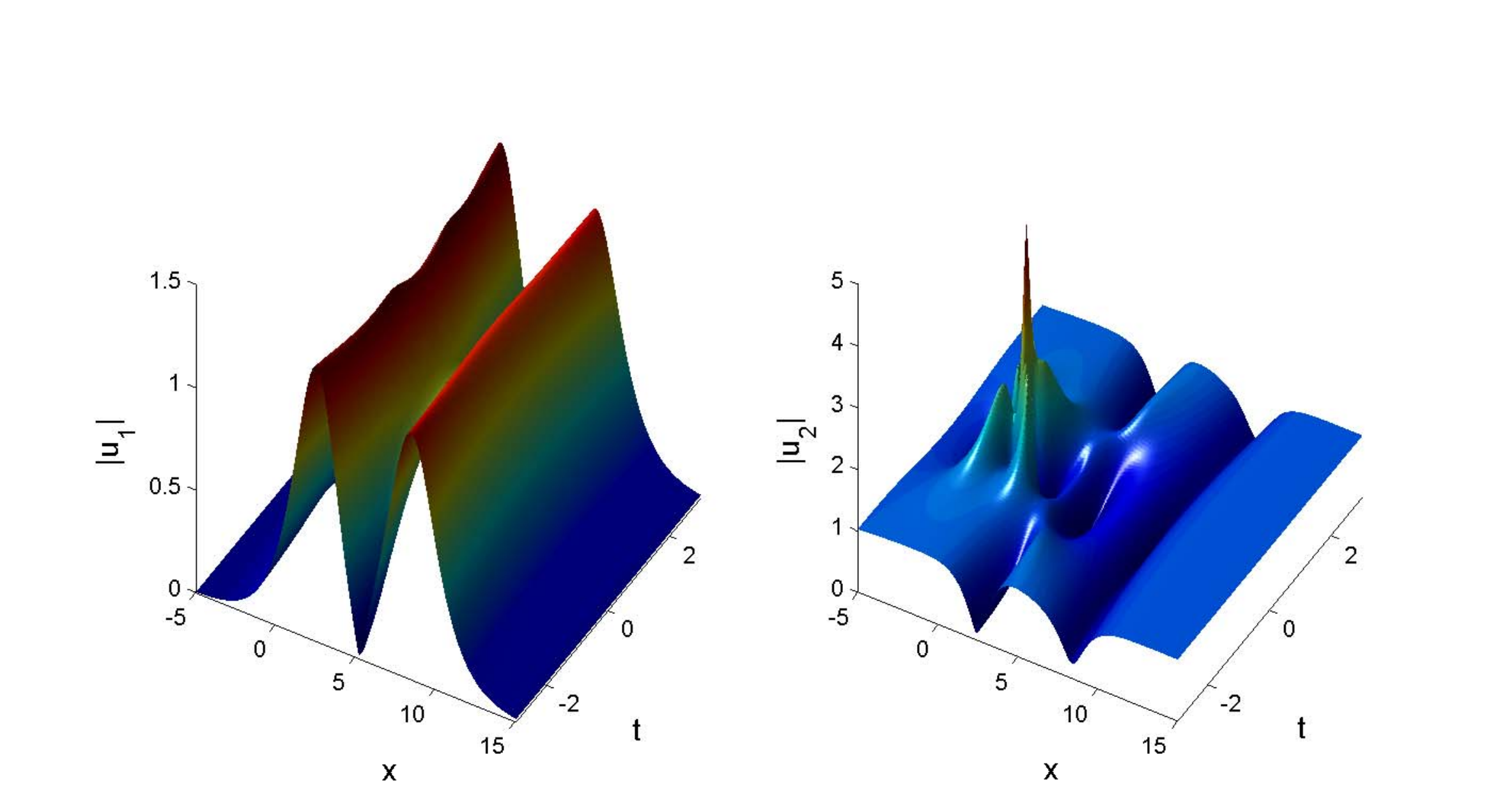}\\
  \caption{ (Color online) Second order rogue waves emerging in the propagation of two bright-dark solitons. The parameters are $a_1=0, a_2=1, \mu_{10}=1, \mu_{20}=1, \mu_{30}=100, \mu_{11}=-1, \mu_{21}=-1$ and $\mu_{31}=-100$.}
   \label{fig:7}
\end{figure*}

\begin{figure*}
  \centering
  \includegraphics[width=5in]{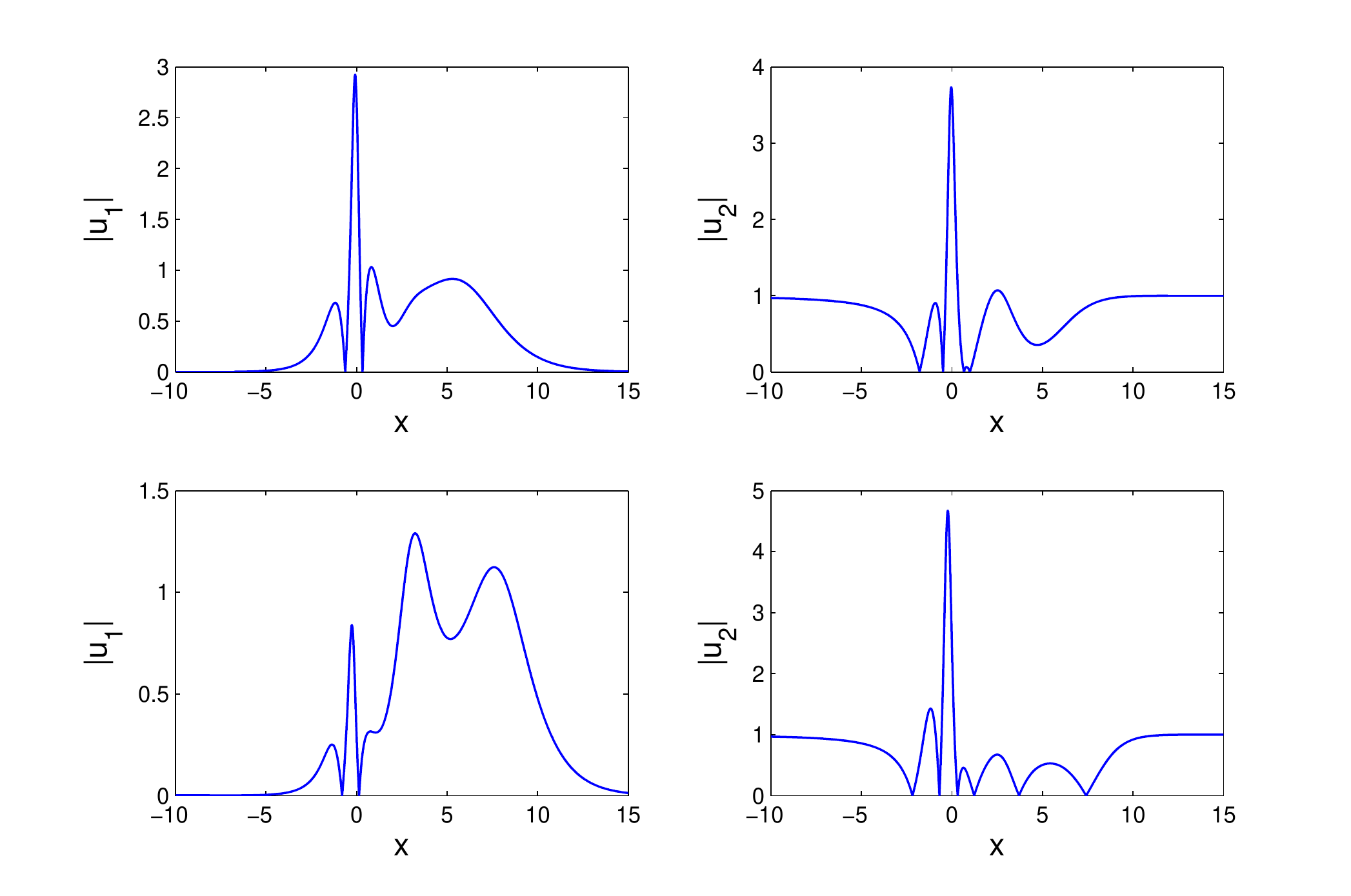}\\
  \caption{ (Color online) Plots of the amplitudes of the functions $u_1(x,0),u_2(x,0)$ in Figures \ref{fig:5}   and \ref{fig:6}.}
   \label{fig:8}
\end{figure*}

\begin{figure*}
  \centering
  \includegraphics[width=5in]{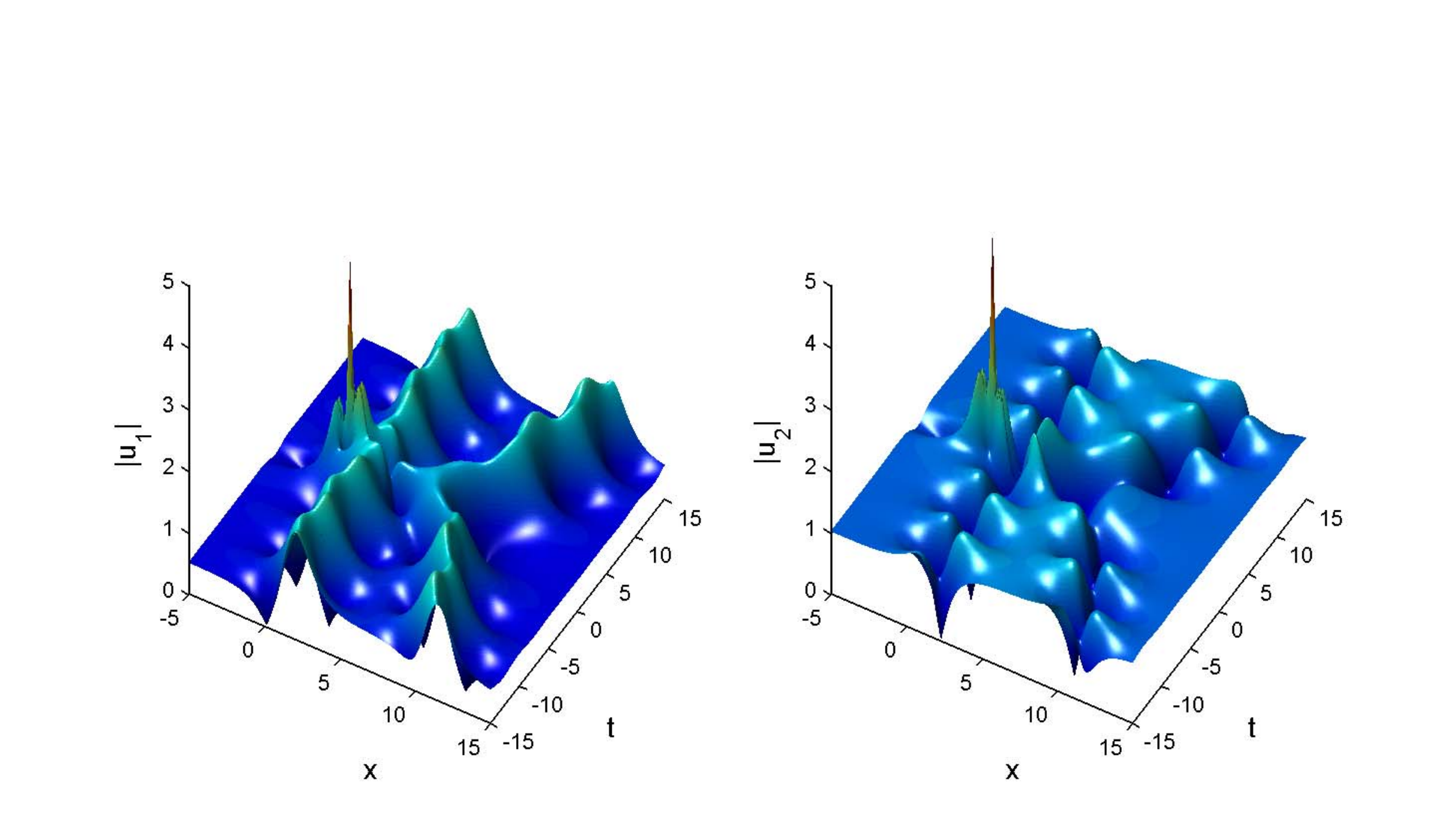}\\
  \caption{ (Color online) Second order rogue waves emerging in the propagation of two breathing bright-dark solitons.
  The parameters are $a_1=\frac{1}{2}, a_2=1, \mu_{10}=1, \mu_{20}=1, \mu_{30}=1, \mu_{11}=0, \mu_{21}=0$ and $\mu_{31}=0$.}
   \label{fig:9}
\end{figure*}

\begin{figure*}
  \centering
  \includegraphics[width=5in]{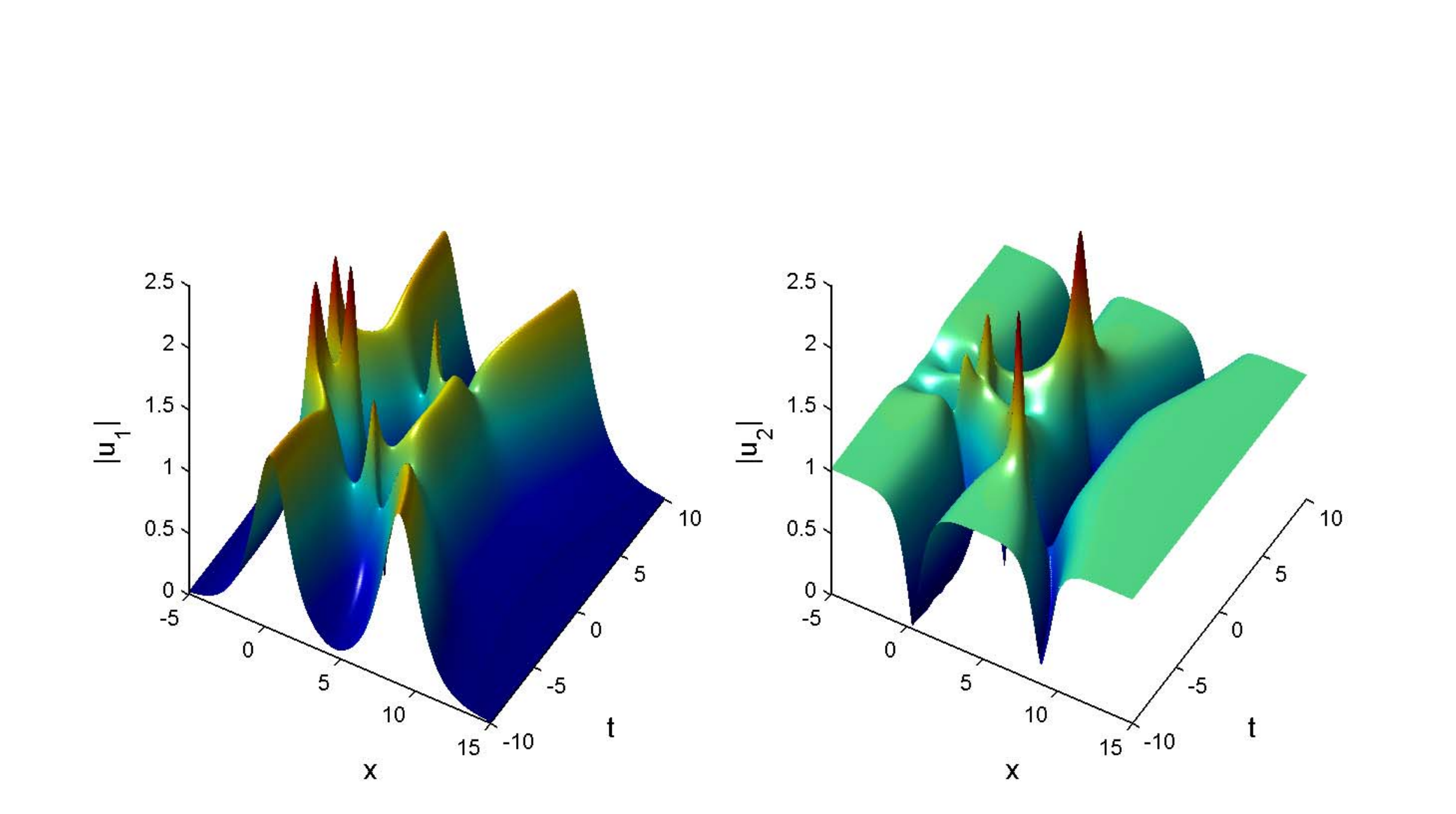}\\
  \caption{ (Color online) Muti-rogue wave patterns emerging in the propagation of two bright-dark solitons. The parameters are $a_1=0, a_2=1, \mu_{10}=1, \mu_{20}=1, \mu_{30}=0, \mu_{11}=1, \mu_{21}=1$ and $\mu_{31}=15$. }
   \label{fig:10}
\end{figure*}

\begin{figure*}
  \centering
  \includegraphics[width=5in]{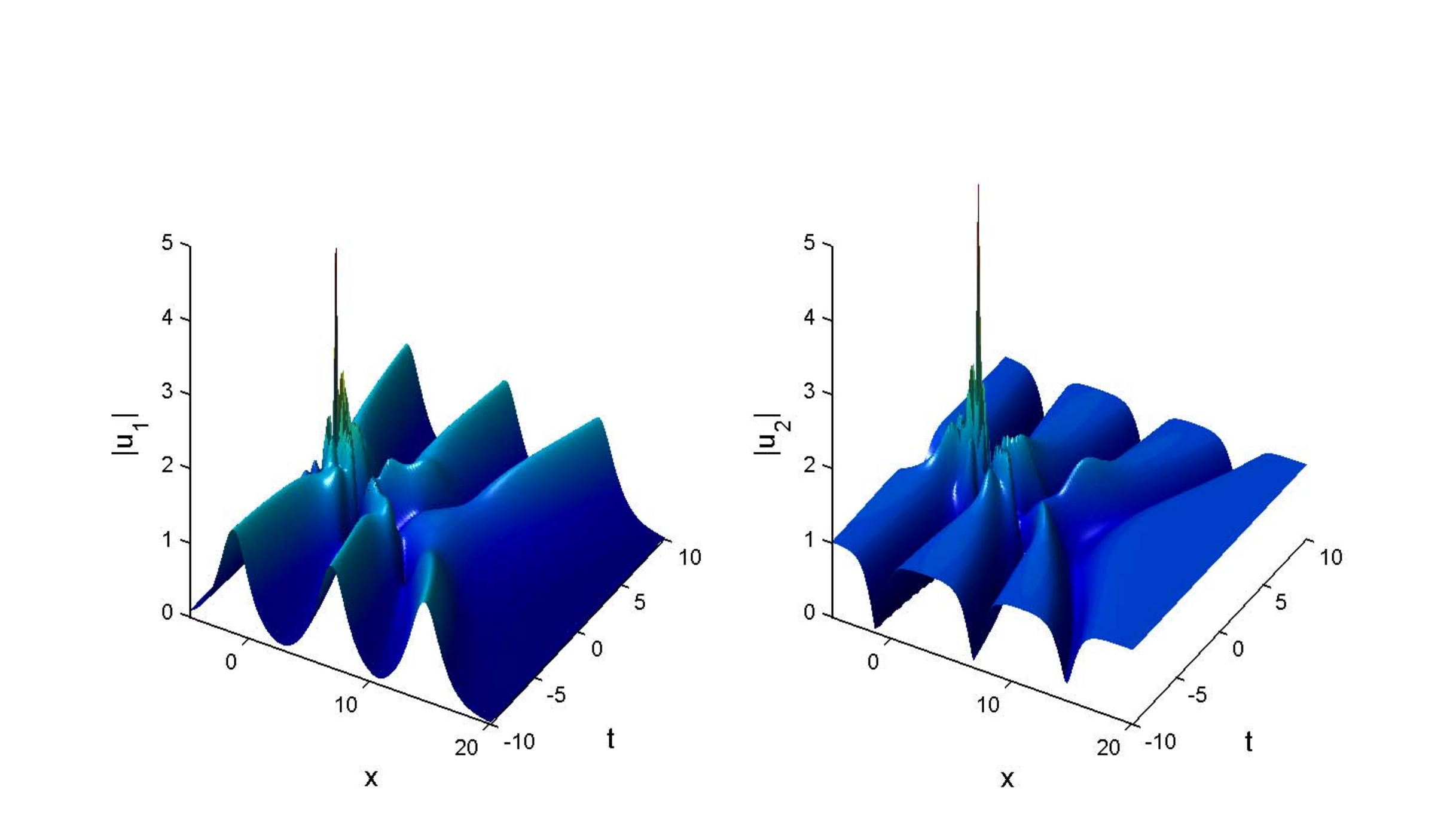}\\
  \caption{ (Color online)  Third order rogue wave patterns emerging in the propagation of three bright-dark solitons. The parameters are $a_1=0, a_2=1, \mu_{10}=1, \mu_{20}=1, \mu_{30}=1, \mu_{11}=0, \mu_{21}=0, \mu_{31}=0, \mu_{12}=0, \mu_{22}=0$ and $\mu_{32}=0$.}
   \label{fig:11}
\end{figure*}

\begin{figure*}
  \centering
  \includegraphics[width=5in]{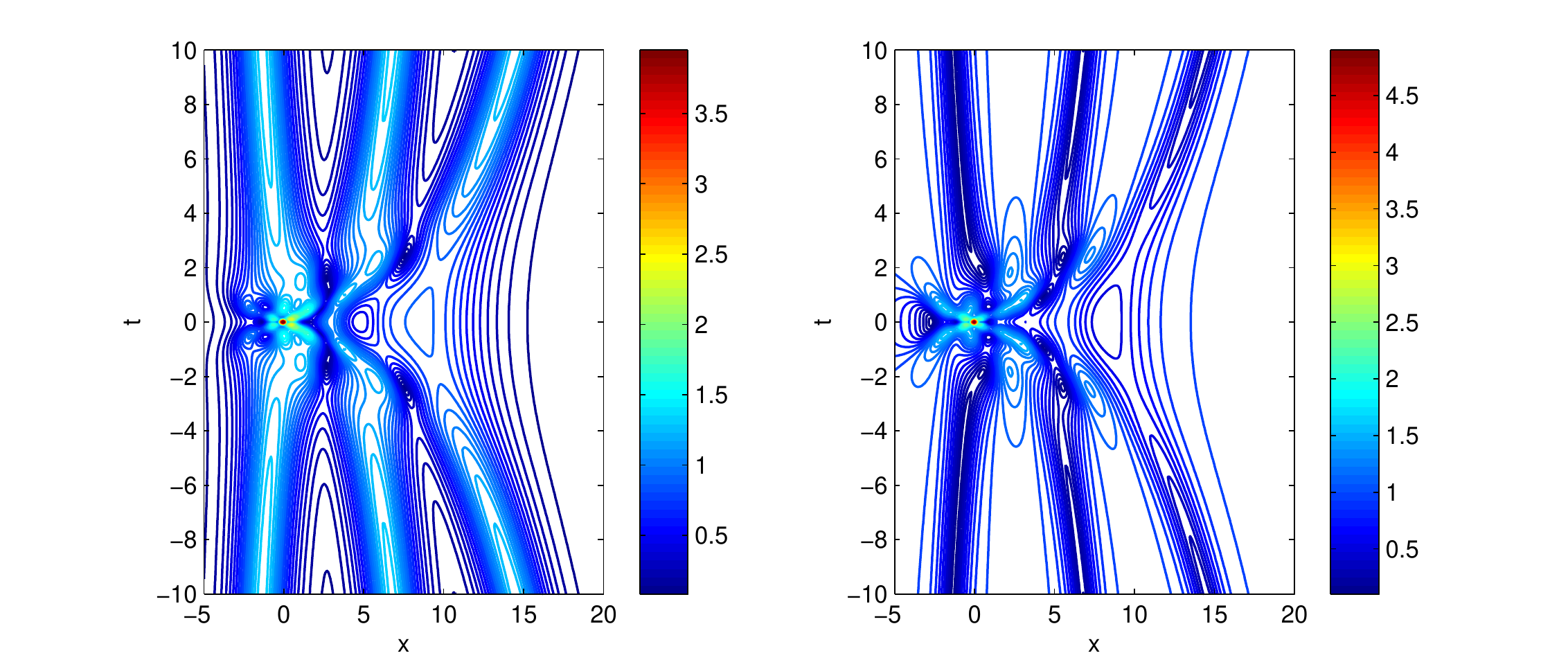}\\
  \caption{ (Color online)  Third order rogue waves emerging in the propagation of three bright-dark solitons. The parameters are $a_1=0, a_2=1, \mu_{10}=1, \mu_{20}=1, \mu_{30}=1, \mu_{11}=0, \mu_{21}=0, \mu_{31}=0, \mu_{12}=0, \mu_{22}=0$ and $\mu_{32}=0$.}
   \label{fig:12}
\end{figure*}

\begin{figure*}
  \centering
  \includegraphics[width=5in]{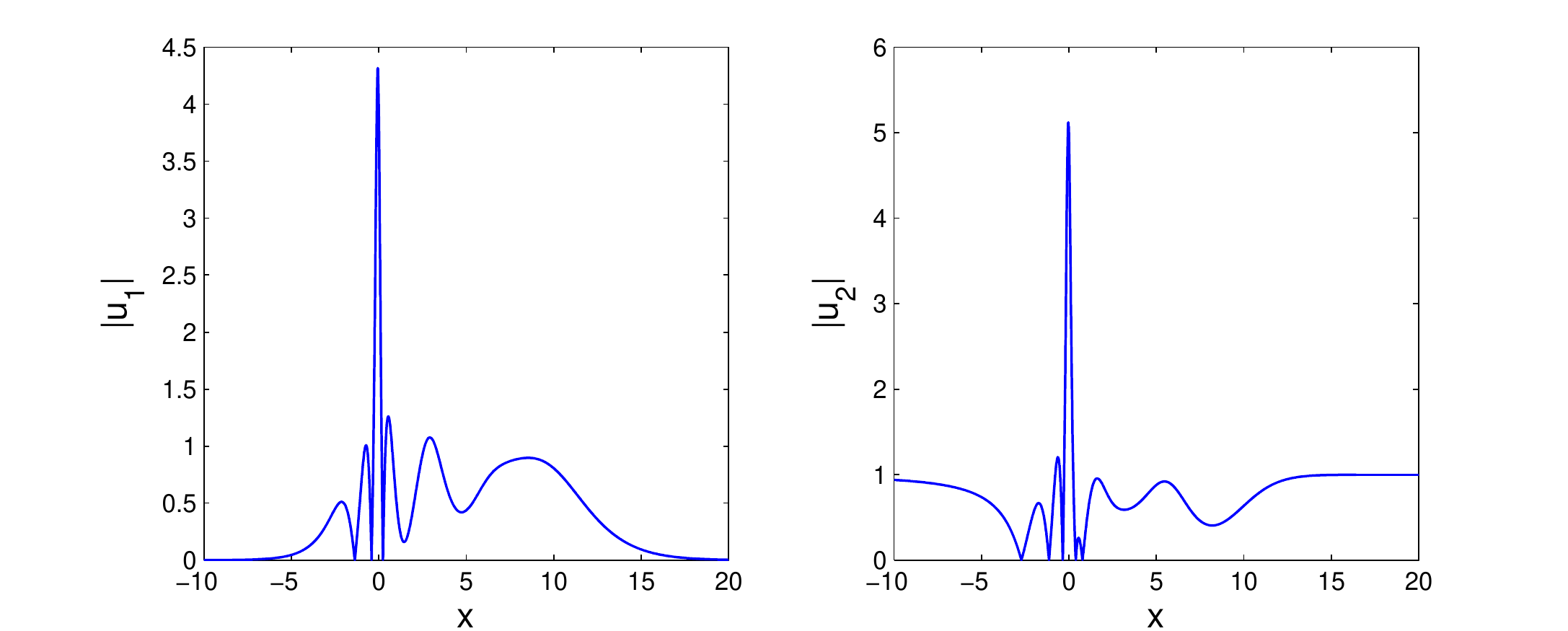}\\
  \caption{ (Color online)  Plots of the amplitudes of the functions $u_1(x,0),u_2(x,0)$ in Figure \ref{fig:11}.}
   \label{fig:13}
\end{figure*}

Figures \ref{fig:6} to \ref{fig:10}  illustrate four different kinds of spatial-temporal distribution patterns
for these  second order solutions (\ref{6.2}) for different parameter values.
In figure \ref{fig:6}  under the condition that the vector $W_1 =0$,  we see that two solitons and a second-order rogue wave coexist.
The second-order rogue wave in $|u_1(x,t)|$ has a peak whose amplitude is about $3$ and  reaches $4$ times
that of the background plane wave amplitude in $|u_2(x,t)|$. In figure \ref{fig:7}, we set $W_1=-W_0$,
and then  the component $u_1(x,t)$ in the second-order rogue wave is difficult to see,  while the other component
$u_2(x,t)$ approaches $5$ times that of the background plane wave amplitude.
This shows  that the amplitudes of the peaks of the second-order rogue wave depend on the values of these free parameters.
This can be observed more clearly in  Figure \ref{fig:8}.
Similarly to the first-order rogue wave, if the parameter $a_1\ne 0 $ ias shown in Figure \ref{fig:9},
both the two bright solitons, and the two dark solitons display breathing behaviour.
Figure \ref{fig:10}  shows that five intensity humps  appear at different times and/or space.
In Figures \ref{fig:11} to \ref{fig:13}  we show some  typical examples of the third-order rogue waves coexisting with three solitons.
These third-order rogue waves have even higher amplitudes, as has been found for the single-component NLSE equation in Ref. \cite{akh},\cite{akha}-\cite{g1},\cite{ked}.
By adjusting the values of the free parameters, more such interesting spatial-temporal patterns can be found.

These results shown above can be extended naturally to fourth-, and even higher-order rogue waves.
We find that  the higher-order rogue waves can be obtained most easily if we only consider the parameter $W_0$.
Moreover, if all the parameters $a_1, a_2, W_i$ are nonzero, the soliton-like waves  perform  breathing behaviour.
In addition, multiple intensity humps appear at different times and/or space provided the vector $W_i$ is not
proportional to the vector $W_j$ for $i\neq j$. Finally, we note that the velocities of  the multisoliton solutions
generally depend on time,  just like the one-soliton case studied by Baronio \emph{et al.} \cite{fa}
The dynamic behaviours found here could be interpreted as a mechanism to excite rogue wave with higher amplitude
out of multiple slowly moving breathing solitons.

\section{Conclusion}

In this paper we have shown how to construct new higher-order rogue waves in the
vector nonlinear Schr\"{o}dinger equations (Manakov system) (\ref{1}) by using a Darboux-dressing transformation
combined with an asymptotic expansion.  These $N$th-order rogue wave solutions contain $3N$ parameters, and can be expressed
explicitly in a separation of variable form.  The found solutions exhibit a range of interesting and complicated dynamics, obtained by varying
the available parameters.  These include  bright-dark higher-order rogue waves, bright-dark multi-rogue waves patterns,  and
rogue waves interacting with multisolitons.  These new spatial-temporal patterns reveal the potential rich dynamics in rogue wave solutions,
and although found here in a VNLSE system, rather than in the usual single-component NLSE,
promote our understanding of rogue-wave phenomena.

Although our solutions exhibited here have set the wavenumber $k =0$,  the same procedure can be used when $k \ne 0$.
 One begins again with the solution (\ref{3.2}) and the following analysis is only slightly more complex.
Further, since in a vector system modulational  instability may occur in the normal dispersion regime resulting from cross-phase modulation,
the corresponding rogue wave solutions are of interest, and the technique presented in this paper is also available for Matrix nonlinear Schr\"{o}dinger equation\cite{qin}, these will be reported elsewhere.
Also, the novel connection between mulitsolitons and rogue waves revealed here has potentially far-reaching significance,
and a topic we will explore in the future.

\section{Acknowledgments}

The work was supported in part by the National Natural Science Foundation of China (No. 10801037), the New Teacher
Grant of Ministry of Education of China (No. 200802461007), the Young Teachers Foundation (No. 1411018) of Fudan
university and Yunnan province project Education Fund: No. 2013C012. Also, the authors are very grateful to Professor Peter
D. Miller, Professor John E. Fornaess and Professor Zhengde Dai for their enthusiastic support and useful suggestions. \\

\end{document}